%% file: iclr2027_conference.tex
\documentclass{article}
\usepackage{iclr2027_conference,times}

\iclrfinalcopy

\AddToHook{cmd/maketitle/after}{\fancyhead[L]{Preprint}}

\input{math_commands.tex}

\usepackage{url}
\usepackage{tikz}
\usepackage{booktabs}
\usepackage{colortbl}
\usepackage{microtype}
\usepackage{wrapfig}
\usepackage{float}
\usepackage{hyperref}
\newcounter{algorithm}
\usepackage{enumitem}
\usetikzlibrary{arrows.meta,positioning}
\input{CONTEXT.tex}
\title{\Title}

\author{
Shuhuai Li \\
Shanghai Jiao Tong University
\And
Shengdong Ye\\
Shanghai Jiao Tong University
\And
Dongdong Ge \\
Shanghai Jiao Tong University
\And
Yinyu Ye \\
Stanford University
\And
Jianghao Lin\thanks{Corresponding author. E-mail: \texttt{linjianghao@sjtu.edu.cn}.} \\
Shanghai Jiao Tong University
}

\begin{document}
\maketitle

\input{sections/abs}

\input{sections/intro}

\input{sections/bg}

\input{sections/method}

\input{sections/experiments.tex}

\input{sections/related-work}


\input{sections/conclusion}

\subsection*{AI use statement}

We used generative AI tools to assist with manuscript editing and wording review. We did not use generative AI to run experiments, fabricate measurements, or decide which experimental artifacts are valid. We reviewed all AI-assisted text and figure suggestions against the source code, experiment records, cited literature, and the compiled manuscript. The authors take responsibility for the final content, claims, and artifacts in this submission.

\bibliographystyle{iclr2027_conference}
\bibliography{iclr2027_conference,references}
\clearpage
\appendix
\input{sections/prediction-appendix.tex}
\input{sections/method-appendix.tex}
\input{sections/async-bound-proof.tex}
\input{sections/glm-traces-appendix.tex}

\section{Compatibility}
\input{sections/tp8-appendix.tex}

\input{sections/tokencake-appendix.tex}
\input{sections/experiments/inferencex-replay.tex}
\input{sections/tool-delay-appendix.tex}

\input{sections/placement-case-studies.tex}

\input{sections/overhead-appendix.tex}

\input{sections/related-work-appendix.tex}
\end{document}

%% file: math_commands.tex
\usepackage{amsmath,amsfonts,bm}

\def\eqref#1{equation~\ref{#1}}

\def\1{\bm{1}}

\def\vrr{{\textnormal{r}}}

\DeclareMathAlphabet{\mathsfit}{\encodingdefault}{\sfdefault}{m}{sl}
\SetMathAlphabet{\mathsfit}{bold}{\encodingdefault}{\sfdefault}{bx}{n}



%% file: CONTEXT.tex
\newcommand{\Ours}{TAO}
\newcommand{\Title}{\Ours: \underline{T}ool-\underline{A}ware J\underline{o}int Scheduling for LLM-Based Agentic Serving}
\newcommand{\Serve}{Agent Session}

\newcommand{\ReadyNext}{readiness}

\newcommand{\Acting}{tool execution}
\newcommand{\ValueAssign}{session evaluation}

\newcommand{\Scene}{CGDC}

\newcommand{\Session}{agent session}
\newcommand{\Sessions}{agent sessions}
\newcommand{\Interrupt}{session interruption}
\newcommand{\Restart}{session restart}

\newcommand{\EvalDecide}{evaluate-then-decide}
\newcommand{\DecisionMaking}{decision making}
\renewcommand{\vrr}{vLLM Scheduler w/ RR}
\newcommand{\vca}{vLLM Scheduler w/ CA}
\newcommand{\ta}{ThunderAgent}
\newcommand{\dy}{Dynamo}
\newcommand{\TTFT}{TTFT}

\newcommand{\SWEVerified}{SWE-verified}
\newcommand{\TokenCake}{TokenCake}
\newcommand{\SWELite}{SWE-Lite}

%% file: sections/abs.tex
\begin{abstract}
Agentic services alternate between model inference on clustered GPUs and external tools on distributed CPUs. In this scenario, existing systems (1) overlook tool-return readiness in service evaluation, wasting limited resources, and (2) use heuristic strategies to decide admission and placement on GPUs, resulting in suboptimal scheduling decisions. To bridge this gap, we present \Ours{}, a tool-aware joint scheduling framework for LLM-based agentic serving. Its value interface combines near-term tool readiness, node-specific cache credit with time-aware decay, and token footprint; a dynamic program jointly selects and places waiting sessions under capacity constraints. Across the 12 reported model--workload--concurrency settings, \Ours{} reduces latency by up to 26.8\%, reduces mean time to first token (\TTFT{}) by up to 36.3\%, and increases decode throughput by up to 29.5\% relative to the best evaluated state-of-the-art baseline for each metric. 
\end{abstract}

%% file: sections/intro.tex
\section{Introduction}
Recent advances in language models have expanded their use beyond basic chatbots to complex agents, making CPU-based tool execution a substantial source of latency alongside GPU-based inference~\citep{llm-core-mimo-scaling-rl, dsec}. Specifically, an agent operates through an \textbf{\textit{\Session{}}} that alternates between LLM inference and external tools, such as compilers and retrievers, to complete multi-step tasks~\citep{anthropic_tooluse, openai_computeruse}. When tools run outside the inference service, a session's demand for GPU computation becomes interruptible. We call this serving scenario \underline{\textbf{C}}lustered-\underline{\textbf{G}}PUs with \underline{\textbf{D}}istributed-\underline{\textbf{C}}PUs (\Scene{}), illustrated in Figure~\ref{fig:ob}(Left): clustered GPUs, i.e., nodes, provide shared inference capacity, while external tool calls introduce delays between LLM requests\footnote{One agent session consists of multiple LLM inference requests (i.e., steps) after tool calls.}. Hence, efficient agentic serving must account for both inference demand and tool timing to sustain throughput and session responsiveness.

We find that existing serving systems generally follow an \textit{\EvalDecide{}} scheduling pattern~\citep{llmd,sglrouter,vllmrouter}. They evaluate node features at the session level and then decide where and when to serve each session. This separates the problem into two related challenges: (1) how to value a session's next model turn, and (2) how to select and place sessions under shared capacity constraints.

\textbf{For \textit{\textbf{\ValueAssign{}}}, existing systems in \Scene{} scenarios generally overlook tool-call latency when evaluating established workload features.} Cache-aware systems reuse KV prefixes and balance inference load~\citep{sglang,preble,mooncake,dynamo}; workflow-aware systems anticipate future calls~\citep{helium,smoothagent}; and latency-aware schedulers prioritize sessions using latency or remaining-work estimates~\citep{elis,kairos}. Tool timing provides another useful signal, yet timing is utilized only as a surrogate for KV reuse or token count: \citet{continuum} predicts tool duration to choose KV-cache retention times, while \citet{thunderagent} discounts the token footprint of sessions executing tools as time elapses. Such features commonly enter fixed scoring rules or manually chosen thresholds.

\textbf{For \textit{\DecisionMaking{}}, existing systems usually apply heuristic scheduling strategies to each session, leading to myopic and suboptimal placements.} In llm-d and SGLang~\citep{llmd,sglrouter}, per-session scoring and routing rules assign inference opportunities \textit{individually}. Yet sessions compete for the same node capacity. In particular, assigning one session can prevent a more valuable combination of other sessions from being served. Furthermore, the heuristic selection and placement rules can also lead to suboptimal choices~\citep{hexagent, kairos_power}. As shown in Figure~\ref{fig:ob}(Right), they fail to formulate scheduling jointly across multiple sessions.

\input{figures/ob}

To address these challenges, we propose \Ours{}, a \underline{T}ool-\underline{A}ware j\underline{O}int scheduling framework for LLM-based agentic serving.
To incorporate \textit{tool latency} into \textbf{\ValueAssign{}}, we propose tool-return readiness and explicitly integrate it with established features such as KV-cache reuse. Following \citet{thunderagent}, we interrupt sessions during tool calls and put them into a \textit{waiting pool}. Tool-return readiness estimates which sessions will need inference within the next scheduling window. For each session--node pair, we form a value triplet from readiness, time-decayed \textit{KV-cache reuse}, and token footprint.
Given these evaluation triplets, we design a \textit{joint scheduling} algorithm in the \textbf{\DecisionMaking{}} stage that considers all sessions in the waiting pool. Across deployments in different scenarios, joint optimization considers how the admission of each session affects the placements available to others. The optimizer maximizes the additive evaluation triplets using dynamic programming. Their coordinates evolve with cache age and tool progress, allowing distinct signals to guide decisions over time without a weighted sum. Separating the optimization procedure from \ValueAssign{} makes the framework extensible.

We summarize our contributions as follows:
\begin{itemize}[leftmargin=1.5em]
    \item We identify the Clustered GPUs with Distributed CPUs (CGDC) serving scenario in the era of LLM-based agents, which shifts the key bottleneck of latency to tool calls.
    \item We propose \Ours{}, a tool-aware joint scheduling framework for LLM-based agentic serving. It improves the evaluate-then-decide paradigm through session evaluation with explicit tool-return readiness and decision making with joint modeling and optimization.
    \item \Ours{} can be generalized and extended to different workloads such as multi-agent orchestration and Looped Agents. Moreover, it is also compatible with non-agent chatbot scenarios.
    \item Across the 12 reported model--workload--concurrency settings, \Ours{} reduces latency by up to 26.8\%, reduces mean time to first token (\TTFT{}) by up to 36.3\%, and increases decode throughput by up to 29.5\% relative to the state-of-the-art baseline for each metric.
\end{itemize}

%% file: figures/ob.tex
\begin{figure}[t]
  \centering
  \includegraphics[width=1\textwidth]{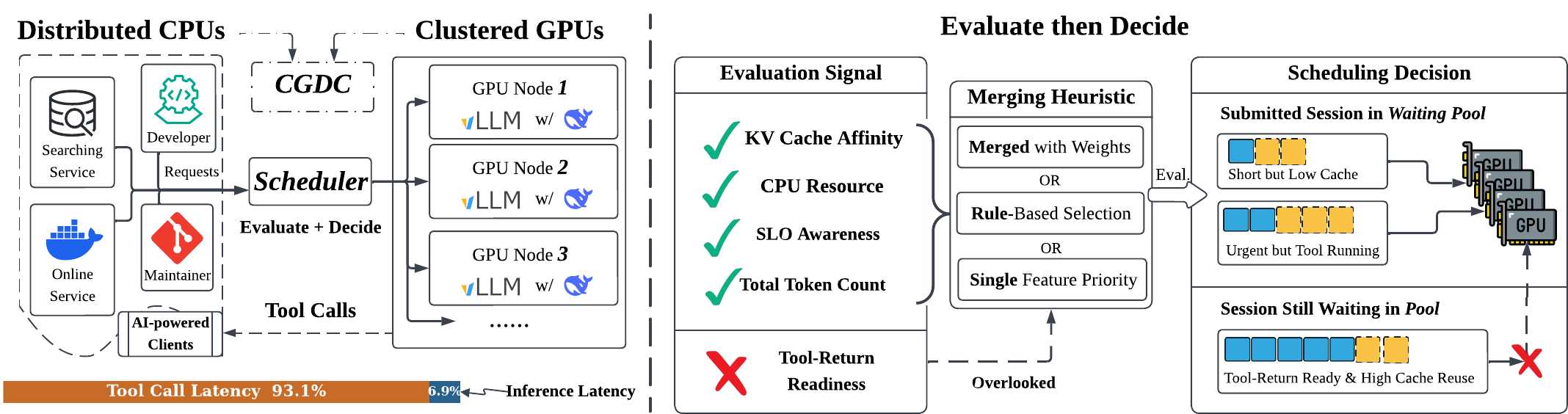}
  \caption{
    Left: In the \Scene{} serving scenario, external tool calls create intervals between LLM requests and opportunities to reallocate inference capacity. Right: Two scheduling challenges motivate \Ours{}, including (1) overlooking tool-return readiness and using heuristic merging for evaluation, and (2) making myopic and suboptimal decisions about session admission and placement.
  }
  \vspace{-2em}
  \label{fig:ob}
\end{figure}

%% file: sections/bg.tex
\section{Background and Problem Description}

\subsection{\Serve{} and distributed resources}
An \Session{} is a persistent interaction that alternates between model requests and external tool calls. During a model request, it uses a serving node; during \Acting{}, it awaits a result outside the inference service. A scheduler records each session's state and controls session admission, while the serving engine manages physical KV-cache residency and eviction.

A later model request can reuse the KV prefix produced by earlier turns and avoid prefill computation~\citep{vllm}. Because an \Session{} carries its instructions, history, and tool observations across turns, reuse depends on both the \Session{} history and the node selected for its next request. SGLang and Preble exploit related prefix structure in model programs and distributed scheduling~\citep{sglang,preble}.

Our \Scene{} separates clustered GPU inference from distributed tool execution. Client or remote tools introduce variable delays between model requests, while their CPU and RAM are managed \textit{outside} the GPU admission budget~\citep{anthropic_tooluse,openai_computeruse,agentsysbench}. Tool latency therefore determines when an \Session{} may need GPU resources for inference again.

\subsection{\Interrupt{} and \Restart{}}
During tool execution, a session can release its GPU resources safely, although its cached prefix may subsequently be evicted. In contrast, preempting an active inference session requires recomputation. Repeated recomputation and reconstruction cause \textit{KV-cache thrashing}, wasting capacity on state that must be rebuilt. This phenomenon motivates cache-aware engine and scheduling work~\citep{vllm,sglang,preble,dynamo}, as well as policies that weigh retaining state against serving other work~\citep{continuum,infercept}.

Therefore, at a tool boundary, the scheduler can apply \textit{\Interrupt{}}. Specifically, a scheduler can interrupt the \Session{}'s model admission and release its node resources while the external tool continues. When inference service is needed again, \textit{session restarting} restores admission on an eligible node and resumes the existing interaction. Session interruption and session restarting describe session state transitions. They are decoupled from the underlying inference engine.

We find that the decisions made for these two operations can be based on the same session evaluation. Thus, we optimize toward a unified goal for maximizing the evaluation.
Session interruption can be decided by ordering sessions according to their values. Interrupted sessions join new arrivals in a waiting pool. We focus on \textit{unified session evaluation} and \textit{restarting decisions} from this pool. 

%% file: sections/method.tex
\section{Method}
\subsection{Framework overview}
We present \Ours{}, which combines agentic \ValueAssign{} with \DecisionMaking, i.e. dynamic programming for joint selection and placement. The detailed formulation appears in Section~\ref{sec:joint-placement}.

Figure~\ref{fig:framework} illustrates the pipeline of our method. First, \ValueAssign{} maps \Session{} and node observations to assignment values (Section~\ref{sec:value-assign}), including KV cache with time-aware decay (Section~\ref{sec:cache-value}) and tool-return readiness (Section~\ref{sec:tool-readiness}). Second, an exact dynamic program jointly selects waiting \Sessions{} and assigns them to nodes under residual capacity constraints (Sections~\ref{sec:joint-placement} and~\ref{sec:exact-dp}). Separating \ValueAssign{} from \DecisionMaking{} lets the evaluation interface evolve while preserving the scheduling procedure.

\input{figures/value-dp-framework.tex}

\subsection{Formulation and Agentic \ValueAssign{}}
\label{sec:joint-placement}
\label{sec:value-assign}
At each restart scheduling decision, let the waiting pool $P$ contain the \Sessions{} awaiting inference admission, including interrupted sessions and new arrivals. Let $B$ be the nodes in the frozen snapshot.
Each \Session{} $i$ has admission cost $w_i$, and each node $b$ has residual capacity $C_b$. Let $E\subseteq P\times B$ denote eligible assignments, and let $x_{i,b}\in\{0,1\}$ indicate whether session $i$ is admitted to node $b$. The feasible set is
\begin{equation}
\label{eq:feasibility}
\mathcal X=\left\{x:
\sum_{b\in B}x_{i,b}\leq 1,\quad
\sum_{i\in P}w_i x_{i,b}\leq C_b,\quad
x_{i,b}=0\ \text{if }(i,b)\notin E
\right\}.
\end{equation}

\Ours{} assigns each \Session{}--node pair the lexicographic value
\begin{equation}
\label{eq:values}
\boldsymbol v_{i,b}=(k_{i,b},r_i,t_i),
\end{equation}
where $k_{i,b}$ is decayed cache credit, $r_i$ is the binary
\ReadyNext{} signal, and $t_i$ is token count. We jointly select and place sessions to maximize
their aggregate value:
\begin{equation}
\label{eq:objective}
x^\star\in
\underset{x\in\mathcal X}{\operatorname{arg\,lexmax}}
\sum_{i\in P}\sum_{b\in B}x_{i,b}\boldsymbol v_{i,b}.
\end{equation}

The evaluation triplet's order remains fixed while its values evolve over time. Our design rationale is detailed as follows:

\begin{itemize}[leftmargin=1.5em]
    \item \textit{KV-Cache credit} takes precedence, but its value decays quickly after interruption, because prefix reuse directly avoids repeated prefill, motivating cache-aware scheduling~\citep{dynamo}. The age-based discount in Section~\ref{sec:cache-value} favors recent cache evidence, which is often most useful when a tool returns quickly.
    \item As unrefreshed credit decays and the first component is equalized, \textit{tool-return readiness} can distinguish \Sessions{} nearing a return from those still far from their next model turn (Section~\ref{sec:tool-readiness}).
    \item \textit{Token footprint} then favors larger accumulated contexts among plans tied on cache credit and readiness. The optimizer accounts for the greater capacity cost of such contexts when choosing the complete set of assignments.
\end{itemize}

\input{figures/cache-decay-evidence.tex}

\subsection{Time-aware cache value for Clustered GPUs}
\label{sec:cache-value}
\Ours{} estimates reusable KV state from observations associated with each target node. It uses an observed cached-token count when available, or estimates the count from observed token volume and cache rate. A node without matching evidence receives zero credit. Figure~\ref{fig:cache-decay-evidence} reports observed cache-token retention in DeepSeek-V4 Flash traces from two nodes with $4\times$H100 per node; the decline in binned medians motivates discounting older observations. Points show binned medians with bootstrap 95\% intervals; the curve is a descriptive exponential fit over the populated bins through 1,800\,s. Older observations receive less credit:
\begin{equation}
\begin{aligned}
&k_{i,b}(t) = k^0_{i,b}\,2^{-\tau_{i,b}/h},\\
&\tau_{i,b} = \max(0,t-t^{\mathrm{obs}}_{i,b}).
\end{aligned}
\end{equation}
Here $k^0_{i,b}$ is the observed cache credit and $h$ is its decay half-life. We use $h=10$ seconds and round the resulting credit for valuation. The discount makes the admission value responsive to observation age; the serving engine continues to control physical allocation, reuse, and eviction. The half-life is a scheduler parameter, distinct from the fitted trace time constant. Dynamo similarly adjusts overlap value under excess prefill load, and ThunderAgent discounts token demand over time~\citep{dynamo,thunderagent}.

\subsection{Tool-return readiness in a distributed CPU}
\label{sec:tool-readiness}
In our \Scene{} deployment, external executors manage tool CPU and RAM, while \Ours{} controls GPU admission. Tool-return readiness connects these resource domains by identifying sessions likely to request model service within a short horizon. Tool timing reflects external execution conditions; the GPU admission budget accounts for the returning session's inference footprint.
Tool interfaces can expose progress or return deadlines: Claude Code's Monitor forwards background-work events, while OpenCode reports termination after a Bash timeout~\citep{claude_monitor,opencode_timeout}.

The predictor combines these hints, parsed sleep commands, and historical tool durations, using direct timing evidence first and falling back to the tool-specific history, global history, or a five-second default~\citep{continuum}. For a call with elapsed time $e_i(t)$, the historical estimate $\widehat D_f$ gives
\begin{equation}
\label{eq:history-duration}
\widehat d_i(t)=\max\{0,\widehat D_f-e_i(t)\}.
\end{equation}
The implementation adds a finite-sample margin to the historical mean; Appendix~\ref{app:prediction} specifies this estimator in detail.

\textbf{Binary readiness.}
Let $o_i(t)$ indicate observed tool completion, $a_i(t)$ indicate an established \Acting{} candidate with an eligible prediction source, and $g_i(t)$ indicate valid prediction timestamps refreshed within 30 seconds. Write $z_i(t)$ for the earlier available predicted completion or exposed return deadline, treating an unavailable deadline as $+\infty$. With horizon $H$ and overdue-probe indicator $b_i(t)$, we define
\begin{equation}
\label{eq:readiness}
r_i(t)=
\begin{cases}
1, & o_i(t)=1,\\
a_i(t)b_i(t), & o_i(t)=0,\ z_i(t)\leq t,\\
a_i(t)g_i(t)\mathbf{1}[z_i(t)\leq t+H], & \text{otherwise}.
\end{cases}
\end{equation}
\Ours{} uses $H=1$ second; the scheduler's \Interrupt{} valuation uses its configured scheduling interval. Overdue predictions receive temporary probe opportunities instead of remaining ready indefinitely. Observed completion takes precedence over prediction checks. This signal prioritizes near-term returns among plans tied on higher-precedence values.

\subsection{Joint scheduling with dynamic programming}
\label{sec:exact-dp}
The optimizer is a capacity-state dynamic program for the multiple-knapsack and generalized-assignment structure~\citep{knapsack,shmoystardos}. Let $\Delta$ be the capacity quantum, $\bar w_i=\lceil w_i/\Delta\rceil$, and $\bar C_b=\lfloor C_b/\Delta\rfloor$. For candidate $j$, let $D_j(\boldsymbol s)$ be the best objective attainable from the first $j$ candidates at the used capacity $\boldsymbol s$, and let $\boldsymbol u_{j,b}$ embed the value and count for assigning $j$ to node $b$:
\begin{equation}
\label{eq:dp}
D_j(\boldsymbol s)=\mathop{\mathrm{lexmax}}\left\{
D_{j-1}(\boldsymbol s),\;
\max_{b\ \mathrm{eligible}}^{\mathrm{lex}}
\left[D_{j-1}(\boldsymbol s-\bar w_j\boldsymbol e_b)
+\boldsymbol u_{j,b}\right]\right\}.
\end{equation}
Only nonnegative states within $\bar{\boldsymbol C}$ are allowed; $D_0(\boldsymbol 0)=\boldsymbol 0$ and other initial states are unreachable. Skipping or assigning each candidate exhausts the feasible choices, so induction gives exactness for the frozen, discretized problem; rounding can exclude plans feasible before discretization.

Node-specific cache observations are decayed and folded into assignment values during $O(nm)$ preprocessing; cache-equivalent nodes can be grouped into $g\leq m$ aggregate capacity dimensions without moving physical KV blocks. The planner runs in an isolated asynchronous worker (detailed in Appendix~\ref{app:method-details}) and revalidates before application; recorded preparation plus completed-solve work is below 0.12\% of runtime in the four-replica sweep (Appendix~\ref{app:overhead-details}). 

%% file: figures/value-dp-framework.tex
\begin{figure}[t]
\centering
\includegraphics[width=1\textwidth]{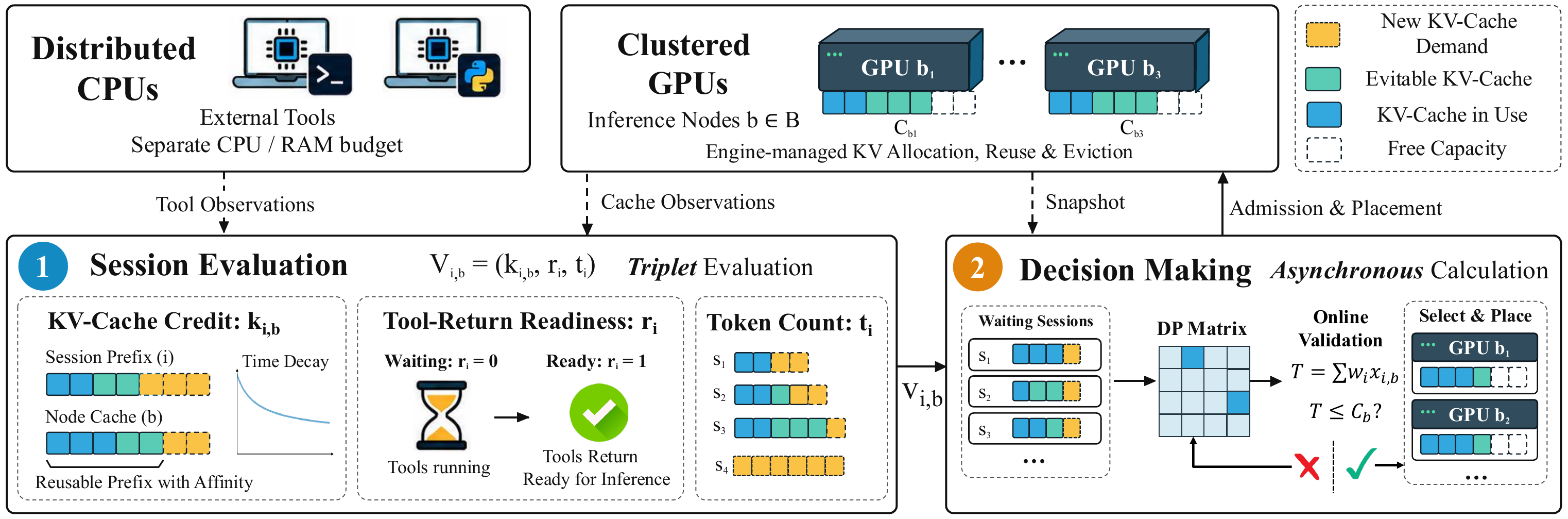}
\caption{The \Ours{} \EvalDecide{} framework in \Scene{}.
External tools on distributed CPUs provide historical tool-call durations, while the serving engines on clustered GPUs provide cache observations.
Then, the \ValueAssign{} component constructs triplets \texttt{(decayed KV-cache credit, binary \ReadyNext{}, token count)}.
The \DecisionMaking{} component jointly selects and places waiting \Sessions{}. The executor validates the plan against live state before applying it.
The resulting assignments schedule sessions to eligible nodes.}
\label{fig:framework}
\end{figure}

%% file: figures/cache-decay-evidence.tex
\begin{wrapfigure}{r}{0.5\textwidth}
    \centering
    \includegraphics[width=\linewidth]{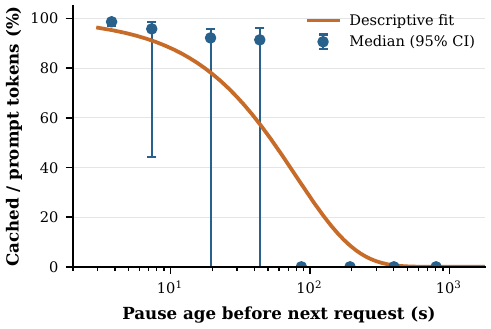}
    \caption{Observed cache-token retention in DeepSeek-V4 Flash traces from two nodes with $4\times$H100 per node.}
    \label{fig:cache-decay-evidence}
    \vspace{-2em}
\end{wrapfigure}

%% file: sections/experiments.tex
\section{Experiments}
\input{tables/main-results.tex}
\label{sec:evaluation}
\input{sections/experiments/setup.tex}

\input{sections/experiments/main-results.tex}
\input{sections/experiments/improvement-breakdown.tex}
\input{sections/experiments/component-ablations.tex}

%% file: tables/main-results.tex
\begin{table}[t]
\centering
\caption{Main results for DeepSeek-V4 Flash on 2 nodes with $4\times$H100 per node and Qwen3-32B on 4 nodes with $2\times$H100 per node. DeepSeek uses Terminal-Bench to study more demanding tasks at moderate concurrency; Qwen3-32B uses \SWEVerified{} at higher concurrency to exercise the capacity of its smaller model replicas. \textbf{Bold} and \underline{underline} mark the best and second-best distinct values in each concurrency-level/metric column.}
\label{tab:main-results}
\scriptsize
\setlength{\tabcolsep}{3.5pt}
\renewcommand{\arraystretch}{1.25}
\resizebox{\linewidth}{!}{%
\begin{tabular}{@{}cl*{9}{r}@{}}
\toprule
\multicolumn{11}{c}{\textit{DeepSeek-V4 Flash, 2 nodes with $4\times$H100 per node}}\\
\midrule
 & & \multicolumn{3}{c}{16 workers} & \multicolumn{3}{c}{32 workers} & \multicolumn{3}{c}{64 workers}\\
\cmidrule(lr){3-5}\cmidrule(lr){6-8}\cmidrule(l){9-11}
 Benchmark & Method & Latency $\downarrow$ & TTFT $\downarrow$ & Decode $\uparrow$ & Latency $\downarrow$ & TTFT $\downarrow$ & Decode $\uparrow$ & Latency $\downarrow$ & TTFT $\downarrow$ & Decode $\uparrow$\\
\midrule
 & \dy{} & 121.993 & \underline{0.146} & 874.085 & 152.297 & \underline{0.176} & \underline{1393.267} & 269.801 & \underline{0.351} & 1548.472\\
 & \vrr{} & 122.814 & 0.177 & 888.489 & 158.841 & 0.217 & 1145.560 & 290.686 & 0.508 & 1299.754\\
 \SWELite{}& \vca{} & \underline{116.011} & 0.183 & \underline{889.590} & \underline{150.431} & 0.226 & 1256.064 & \underline{253.228} & 0.410 & \underline{1561.307}\\
 & \ta{} & 123.564 & 0.379 & 796.215 & 215.727 & 2.708 & 961.791 & 407.180 & 7.447 & 969.018\\
\rowcolor{blue!8}
 & \Ours{} (Ours) & \textbf{113.264} & \textbf{0.135} & \textbf{934.967} & \textbf{142.332} & \textbf{0.153} & \textbf{1477.463} & \textbf{185.410} & \textbf{0.224} & \textbf{2021.523}\\
\rowcolor{gray!15}
 & \textit{Rel. Imprv. (\%)} & +2.4 & +7.4 & +5.1 & +5.4 & +13.3 & +6.0 & +26.8 & +36.3 & +29.5\\
\midrule
 & \dy{} & 536.600 & 1.371 & 494.866 & 579.650 & 1.252 & 651.151 & 688.531 & 1.315 & 760.736\\
 & \vrr{} & 342.464 & \underline{0.306} & \underline{820.754} & \underline{375.182} & 0.499 & 1071.063 & 467.410 & 0.582 & \underline{1597.497}\\
 Terminal-Bench& \vca{} & \underline{304.802} & 0.319 & 723.889 & 394.677 & \underline{0.331} & \underline{1146.434} & 393.065 & \underline{0.387} & \textbf{1654.108}\\
 & \ta{} & 361.061 & 0.336 & \textbf{876.138} & 380.108 & 1.402 & 1053.042 & \underline{391.580} & 1.010 & 1376.539\\
\rowcolor{blue!8}
 & \Ours{} (Ours)& \textbf{298.224} & \textbf{0.239} & 798.624 & \textbf{333.587} & \textbf{0.264} & \textbf{1161.438} & \textbf{376.647} & \textbf{0.324} & 1551.387\\
\rowcolor{gray!15}
 & \textit{Rel. Imprv. (\%)} & +2.2 & +21.7 & -8.8 & +11.1 & +20.3 & +1.3 & +3.8 & +16.2 & -6.2\\
\midrule
\multicolumn{11}{c}{\textit{Qwen3-32B, 4 nodes with $2\times$H100 per node}}\\
\midrule

 & & \multicolumn{3}{c}{96 workers} & \multicolumn{3}{c}{128 workers} & \multicolumn{3}{c}{160 workers}\\
\cmidrule(lr){3-5}\cmidrule(lr){6-8}\cmidrule(l){9-11}
 Benchmark & Method & Latency $\downarrow$ & TTFT $\downarrow$ & Decode $\uparrow$ & Latency $\downarrow$ & TTFT $\downarrow$ & Decode $\uparrow$ & Latency $\downarrow$ & TTFT $\downarrow$ & Decode $\uparrow$\\
\midrule
 & \dy{} & 821.558 & 2.098 & \underline{2074.598} & 857.418 & \underline{7.924} & 1203.708 & 1203.712 & 17.208 & 1239.372\\
 & \vrr{} & 775.303 & 5.109 & 1400.577 & 975.382 & 11.179 & 1359.789 & 1088.036 & 16.645 & 1433.349\\
 \SWELite{}& \vca{} & 811.604 & 5.339 & 1614.671 & 881.068 & 11.707 & 1456.411 & 1179.340 & 18.304 & 1285.413\\
 & \ta{} & \underline{716.272} & \underline{2.051} & \textbf{2104.896} & \underline{839.274} & 8.705 & \underline{2036.280} & \textbf{1028.478} & \underline{16.209} & \underline{1712.363}\\
\rowcolor{blue!8}
 & \Ours{} (Ours)& \textbf{634.694} & \textbf{1.827} & 1996.793 & \textbf{829.361} & \textbf{7.907} & \textbf{2051.807} & \underline{1030.973} & \textbf{14.554} & \textbf{1827.561}\\
\rowcolor{gray!15}
 & \textit{Rel. Imprv. (\%)} & +11.4 & +10.9 & -5.1 & +1.2 & +0.2 & +0.8 & -0.2 & +10.2 & +6.7\\
\midrule
 & \dy{} & 736.823 & \underline{1.985} & 1862.845 & 1041.911 & \underline{10.441} & 1820.919 & 1286.167 & 22.469 & 1667.794\\
 & \vrr{} & 777.584 & 5.082 & 1931.333 & \underline{933.347} & 12.586 & 1804.505 & 1244.525 & 24.221 & 1091.002\\
 \SWEVerified{}& \vca{} & 816.201 & 5.731 & 1522.641 & 1016.513 & 14.263 & 1670.856 & 1264.406 & 23.199 & 1619.768\\
 & \ta{} & \underline{699.801} & 2.396 & \underline{2184.933} & \textbf{911.097} & 10.500 & \underline{2045.875} & \underline{1202.215} & \underline{21.608} & \underline{1993.620}\\
\rowcolor{blue!8}
 & \Ours{} (Ours)& \textbf{642.017} & \textbf{1.392} & \textbf{2354.709} & 933.678 & \textbf{9.092} & \textbf{2128.599} & \textbf{1094.318} & \textbf{17.619} & \textbf{2087.191}\\
\rowcolor{gray!15}
 & \textit{Rel. Imprv. (\%)} & +8.3 & +29.9 & +7.8 & -2.5 & +12.9 & +4.0 & +9.0 & +18.5 & +4.7\\
\bottomrule
\end{tabular}
}
\end{table}

%% file: sections/experiments/setup.tex
\subsection{Experiment Setup}
\textbf{Models and workloads.} We evaluate DeepSeek-V4 Flash, Qwen3-32B, and GLM-4.7~\citep{deepseek, qwen3, glm47} on the SWE-bench Lite workload (\SWELite{}) using the mini-SWE-agent harness (MiniSWE)~\citep{miniswe,swebench}. For DeepSeek, we also use Claude Code on Terminal-Bench 2.0~\citep{terminalbench} for challenging tasks. For Qwen, we test with \SWEVerified{} to exercise the serving capacity of its smaller model replicas. We further evaluate GLM-4.7 on \SWELite{} to test single-node performance. Appendix~\ref{sec:inferencex-replay} reports an InferenceX trace-replay evaluation.

\textbf{Deployment.} DeepSeek runs on 2 nodes with $4\times$H100 per node; Qwen runs on 4 nodes with $2\times$H100 per node. Each node hosts one tensor-parallel model replica. GLM-4.7 runs on one node with $8\times$H100. Appendix~\ref{app:tp8-miniswe} reports a lower-concurrency DeepSeek check on the same single-node topology.

\textbf{Baselines.} \vrr{} and \vca{} denote the vLLM schedulers with round-robin and cache-aware policies, respectively~\citep{vllmrouter}, which align with the policy in SGLang~\citep{sglang}. These baselines cover load spreading and prefix-locality scheduling. \dy{} is a KV-aware scheduler~\citep{dynamo}. \ta{} uses token-count scheduling and best-fit-decreasing placement~\citep{thunderagent}. \Ours{} combines predictive cache-aware session evaluation with joint dynamic-programming placement.

\textbf{Metrics.} We report mean task end-to-end latency, mean time to first token (\TTFT{}), and decode throughput in tokens per second (TPS). Task latency includes both model inference and tool execution. Component analyses also measure prefix-cache hit rate, tool-end-to-first-token delay, and pause wait, which includes the tool call latency of an interrupted session.

%% file: sections/experiments/main-results.tex
\subsection{Main Results}
\label{sec:main-results}
Table~\ref{tab:main-results} compares five baselines across different levels of concurrency, models, and topologies. Taken together, these results make \Ours{} the strongest overall choice: it combines consistently low latency and \TTFT{} with high decode throughput more effectively than any single baseline, consistent with our focus on responsiveness after tool calls and improved KV reuse.

\paragraph{DeepSeek-V4 Flash.} We evaluate on challenging benchmarks with moderate concurrency. On \SWELite{}, , relative to the best competing method in each column, latency falls by 2.4--26.8\%, \TTFT{} by 7.4--36.3\%, and decode throughput rises by 5.1--29.5\%. On Terminal-Bench, \Ours{} also leads on latency and \TTFT{} at all three concurrency levels. 

\paragraph{Qwen3-32B.} We use higher concurrency to exercise the serving capacity of its smaller model replicas. On \SWELite{}, \Ours{} achieves the lowest \TTFT{} at all three worker counts and the highest decode throughput at 96, 128 and 160 workers. On \SWEVerified{}, \Ours{} leads on \TTFT{} and decode throughput throughout, with reductions of 12.9--29.9\% in \TTFT{} and gains of 4.0--7.8\% in decode throughput. It also reduces latency by 8.3\% and 9.0\% at 96 and 160 workers; at 128 workers, latency is 2.5\% higher than ThunderAgent.

\paragraph{GLM-4.7.} As a larger-model setting, GLM-4.7 provides a high-concurrency extensibility showcase on one node with $8\times$H100. Table~\ref{tab:glm47-high-concurrency} reports the separate worker-grouped results: on \SWELite{} at 160 workers, improvements reach 15.5\% in latency, 14.05\% in \TTFT{}, and 20.48\% in decode throughput. On all 500 \SWEVerified{} tasks, \Ours{} reduces mean task latency by 16.2--18.9\% and client \TTFT{} by 15.7--38.4\%, while increasing decode throughput by 22.8--24.7\% relative to \ta{}, the SOTA baseline on each metric.

\paragraph{Source of Improvement.} Figure~\ref{fig:glm-traces} illustrates the serving dynamics at 160 workers. During the middle of the serving window, \Ours{} processes ongoing requests more efficiently and steadily admits additional sessions under limited capacity, avoiding sharp bursts in concurrency. Appendix~\ref{app:glm-traces} provides the corresponding traces at other workers.

\input{tables/glm47-high-concurrency.tex}

\input{figures/glm-traces.tex}

\input{sections/experiments/tokencake-compatibility.tex}


%% file: tables/glm47-high-concurrency.tex
\begin{table}[t]
\centering
    \caption{High-concurrency extensibility showcase for a larger model: GLM-4.7 on \SWELite{} and \SWEVerified{} through MiniSWE, one node with $8\times$H100 (TP8), and 300 tasks per run.}
\label{tab:glm47-high-concurrency}
\scriptsize
\setlength{\tabcolsep}{3.5pt}
\renewcommand{\arraystretch}{1.25}
\resizebox{\linewidth}{!}{%
\begin{tabular}{@{}cl*{9}{r}@{}}
\toprule
\multicolumn{11}{c}{\textit{GLM-4.7, one node with $8\times$H100 (TP8)}}\\
\midrule
 & & \multicolumn{3}{c}{96 workers} & \multicolumn{3}{c}{128 workers} & \multicolumn{3}{c}{160 workers}\\
\cmidrule(lr){3-5}\cmidrule(lr){6-8}\cmidrule(l){9-11}
Benchmark & Method & Latency $\downarrow$ & TTFT $\downarrow$ & Decode $\uparrow$ & Latency $\downarrow$ & TTFT $\downarrow$ & Decode $\uparrow$ & Latency $\downarrow$ & TTFT $\downarrow$ & Decode $\uparrow$\\
\midrule
 & \dy{} & 2870.313 & 21.677 & 290.532 & 3779.196 & 41.865 & 287.844 & 4786.720 & 59.548 & 262.215 \\
 & \vrr{} & 2687.766 & 20.927 & 310.800 & 3633.795 & 39.502 & 294.582 & 4390.961 & 53.588 & 272.460 \\
\SWELite{} & \vca{} & 2606.523 & 19.553 & 318.687 & 3435.779 & 37.665 & 307.983 & 4314.458 & 56.326 & 297.280 \\
 & \ta{} & \underline{999.624} & \underline{7.014} & \underline{773.250} & \underline{1360.382} & \underline{14.243} & \underline{731.052} & \underline{1747.028} & \underline{23.345} & \underline{668.940} \\
\rowcolor{blue!8}
 & \Ours{} (Ours)& \textbf{887.663} & \textbf{6.777} & \textbf{869.628} & \textbf{1261.906} & \textbf{14.103} & \textbf{813.653} & \textbf{1476.190} & \textbf{20.066} & \textbf{805.939} \\
\rowcolor{gray!15}
 & \textit{Rel. Imprv. (\%)} & +11.2 & +3.4 & +12.5 & +7.2 & +1.0 & +11.3 & +15.5 & +14.1 & +20.5 \\
\midrule
  & \dy{} & 3164.578 & 26.761 & 286.215 & 4545.301 & 52.588 & 256.606 & 8217.092 & 86.008 & 163.572 \\
  & \vrr{} & 2938.216 & 23.967 & 292.731 & 4136.027 & 47.930 & 269.838 & 6149.466 & 71.452 & 214.622 \\
 \SWEVerified{} & \vca{} & 2928.613 & 24.505 & 300.339 & 3935.660 & 46.963 & 284.906 & 5989.491 & 71.239 & 220.933 \\
  & \ta{} & \underline{1146.121} & \underline{12.590} & \underline{691.506} & \underline{1533.286} & \underline{18.516} & \underline{688.310} & \underline{1846.499} & \underline{26.045} & \underline{683.551} \\
\rowcolor{blue!8}
  & \Ours{} (Ours) & \textbf{951.720} & \textbf{7.752} & \textbf{856.764} & \textbf{1243.886} & \textbf{15.281} & \textbf{858.140} & \textbf{1548.253} & \textbf{21.943} & \textbf{839.149} \\
\rowcolor{gray!15}
 & \textit{Rel. Imprv. (\%)} & +17.0 & +38.4 & +23.9 & +18.9 & +17.5 & +24.7 & +16.2 & +15.7 & +22.8 \\
\bottomrule
\end{tabular}
}
\end{table}

%% file: figures/glm-traces.tex
\begin{figure}[!h]
\centering
\includegraphics[width=\linewidth]{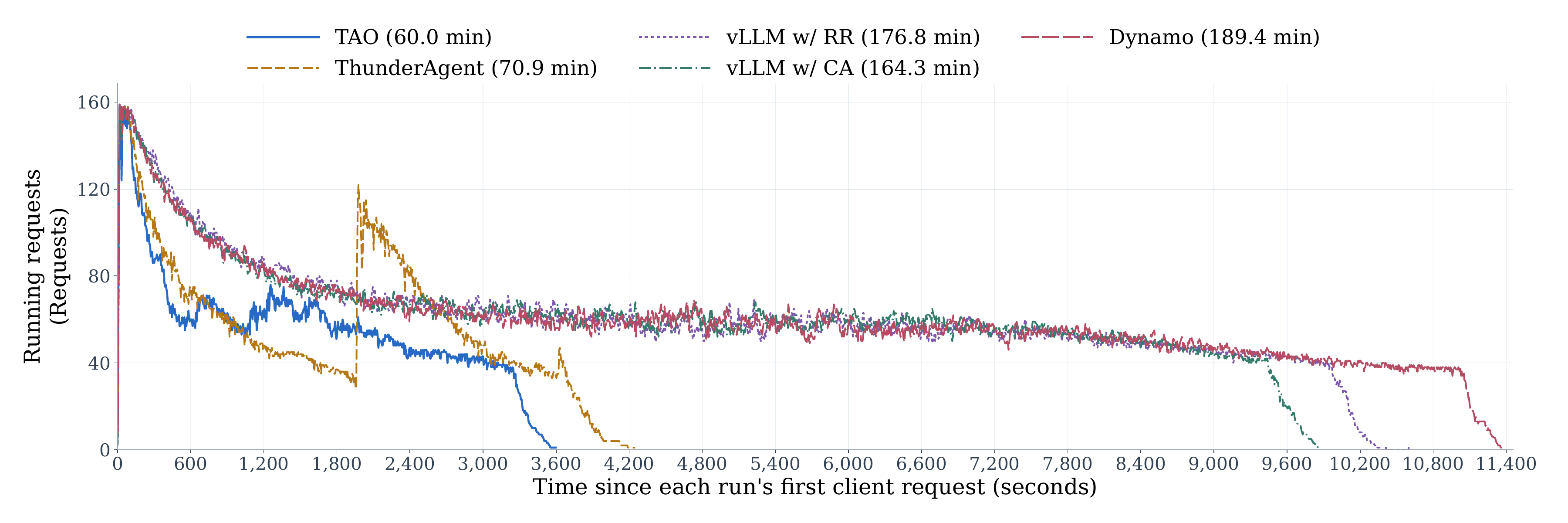}
\caption{Running-request traces for GLM-4.7 on \SWELite{} at 160 workers (TP8). Each run starts at its first client request. \Ours{} completes the workload sooner while sustaining fewer concurrent running requests for much of the run.}
\label{fig:glm-traces}
\end{figure}

%% file: sections/experiments/tokencake-compatibility.tex
\paragraph{Compatibility.} \Ours{} supports heterogeneous agent workloads and composes with engine-level KV-cache management. Table~\ref{tab:mixed-workloads-pending} evaluates a mixture of Terminal-Bench and \SWELite{} alongside AutomationBench at 32, 64, and 128 workers, reporting the available results. Appendix~\ref{app:tokencake} demonstrates integration with \TokenCake{}~\citep{tokencake} and discusses the relationship to program-aware scheduling in Agentix~\citep{agentix} and workflow-aware cache management. Appendix~\ref{app:tp8-miniswe} provides the complementary low-concurrency DeepSeek evaluation on a single TP8 replica.
\input{tables/mixed-workloads-placeholder.tex}

%% file: tables/mixed-workloads-placeholder.tex
\begin{table}[t!]
\centering
\caption{Workload compatibility on mixed workloads with chatbot scenarios. Deployed on DeepSeek-V4 Flash with 2 nodes 4$\times$H100. Latency and TTFT are in seconds, and decode throughput is in token/s.}
\label{tab:mixed-workloads-pending}
\scriptsize
\setlength{\tabcolsep}{3.5pt}
\renewcommand{\arraystretch}{1.2}
\resizebox{\linewidth}{!}{%
\begin{tabular}{@{}l*{9}{r}@{}}
\toprule
 & \multicolumn{3}{c}{32 workers} & \multicolumn{3}{c}{64 workers} & \multicolumn{3}{c}{128 workers}\\
\cmidrule(lr){2-4}\cmidrule(lr){5-7}\cmidrule(l){8-10}
Method & Latency $\downarrow$ & TTFT $\downarrow$ & Decode $\uparrow$ & Latency $\downarrow$ & TTFT $\downarrow$ & Decode $\uparrow$ & Latency $\downarrow$ & TTFT $\downarrow$ & Decode $\uparrow$\\
\midrule
\multicolumn{10}{c}{Terminal-Bench + \SWELite{} (389 tasks)}\\
\midrule
\dy{} & 378.542 & \textbf{0.524} & 897.840 & 575.024 & \underline{0.786} & 949.396 & 1065.101 & \underline{1.331} & 723.171\\
\vrr{} & 419.926 & 1.088 & 832.544 & 866.508 & 3.511 & 681.442 & 1328.289 & 4.823 & 635.044\\
\vca{} & 373.900 & 0.707 & 853.498 & \underline{519.631} & 0.911 & 879.750 & 1002.259 & 1.568 & 751.131\\
\ta{} & \underline{361.778} & 1.149 & \underline{908.751} & 535.400 & 4.200 & \underline{1053.911} & \underline{679.212} & 7.235 & \underline{1337.560}\\
\rowcolor{blue!8}
\Ours{} (Ours) & \textbf{317.476}$^*$ & \underline{0.553}$^*$ & \textbf{1019.827}$^*$ & \textbf{398.911} & \textbf{0.609} & \textbf{1144.432} & \textbf{585.804} & \textbf{0.840} & \textbf{1396.591}\\
\rowcolor{gray!15}
\textit{Rel. Imprv. (\%)} & +12.2 & -5.5 & +12.2 & +23.2 & +22.5 & +8.6 & +13.8 & +36.9 & +4.4\\
\midrule
\multicolumn{10}{c}{AutomationBench (800 tasks)}\\
\midrule
\dy{} & 147.378 & \underline{0.330} & 1174.370 & 593.958 & 1.449 & 574.021 & 1417.212 & \textbf{3.166} & 482.381\\
\vrr{} & 293.286 & 1.849 & 555.398 & 691.139 & 3.616 & 468.963 & 1232.168 & 10.884 & 496.417\\
\vca{} & 129.716 & 0.409 & 1283.423 & 312.150 & \underline{1.046} & 988.401 & {1214.511} & \underline{3.938} & 501.852\\
\ta{} & \underline{107.485} & 0.444 & \underline{1487.097} & \underline{237.577} & 5.368 & \underline{1264.527} & \textbf{523.676} & 16.870 & \textbf{1140.862}\\
\rowcolor{blue!8}
\Ours{} (Ours) & \textbf{101.064} & \textbf{0.246} & \textbf{1604.639} & \textbf{197.023} & \textbf{0.539} & \textbf{1409.613} & \underline{563.114} & 19.123 & \underline{1013.157}\\
\rowcolor{gray!15}
\textit{Rel. Imprv. (\%)} & +6.0 & +25.6 & +7.9 & +17.1 & +48.5 & +11.5 & -7.0 & -504.0 & -11.2\\
\bottomrule
\end{tabular}
}
\end{table}


%% file: sections/experiments/improvement-breakdown.tex
\subsection{Improvement Breakdown}
\label{sec:improvement-breakdown}
As shown in Section~\ref{sec:main-results}, the worker sweep exposes a latency--throughput tradeoff: the lowest latency and \TTFT{} occur at the smallest worker count, while peak decode throughput occurs at 64 workers for DeepSeek \SWELite{} and Terminal-Bench, 128 workers for Qwen \SWELite{}, and 96 workers for \SWEVerified{}. We use DeepSeek at 32 workers as a balanced operating point and examine how cache-aware valuation, tool readiness, and joint placement improve reuse, tool-return responsiveness, and admission decisions.

\paragraph{Cache-aware valuation.} In a fixed-workload replay of 218 tasks with DP placement held fixed for the \ta{} control and cache-aware variants, cache-token ordering raises prefix-cache hit rate by 0.19 percentage points and reduces mean pause wait by 4.5\% relative to \ta{}. Adding cache-credit decay raises hit rate by 0.31 percentage points and decode throughput by 1.68\% relative to \ta{}. Figure~\ref{fig:tp4x2-gain-breakdown} summarizes these changes; bars are percentage points for cache hit and percentages for the other metrics, with positive values denoting improvement.

\input{figures/tp4x2-gain-breakdown/tp4x2-cache-breakdown.tex}
\paragraph{Tool readiness.} Enabling \ReadyNext{} in the paired \SWELite{} runs increases decode throughput by 4.03\% and reduces tool-end-to-first-token delay by 10.88\%. Mean pause wait falls from 1.067 to 0.972\,s. Figure~\ref{fig:readiness-effectiveness} shows the return-delay distributions and prediction error. Across 10,967 completed tools, the final duration prediction has a mean absolute error of 0.410\,s, with 82.9\% of errors within 0.5\,s.
\input{figures/readiness-effectiveness.tex}
The return-delay CDFs show that readiness shifts model responses earlier after
tool completion. The middle panel summarizes the serving gains, and the right
panel shows the prediction accuracy that informs readiness valuation.
The return CDFs contain 10,481 and 10,673 observations, and the prediction
panel contains 10,967 predictions.

\paragraph{Joint Scheduling.} Joint optimization improves the lexicographic
scheduling objective in six of the 23 replay snapshots with multiple
candidates (26.1\%), matching the short-first heuristic in the remaining 17.
Across all 35 reconstructable snapshots, including single-candidate cases,
DP improves six (17.1\%) and matches the heuristic in 29 (82.9\%).
Detailed cache-credit and admission examples appear in Appendix~\ref{app:placement-cases}.


%% file: figures/tp4x2-gain-breakdown/tp4x2-cache-breakdown.tex
\begin{wrapfigure}{r}{0.5\textwidth}
  \centering
  \includegraphics[width=0.48\textwidth]
      {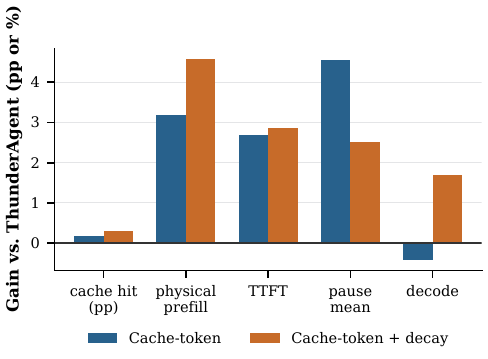}
    \caption{Cache-aware valuation in the 32-worker replay, relative to the \ta{} control with DP placement.
    Reductions in prefill, \TTFT{}, and pause wait are shown as positive improvements.}
    \vspace{-2em}
    \label{fig:tp4x2-gain-breakdown}
\end{wrapfigure}

%% file: figures/readiness-effectiveness.tex
\begin{figure}[t]
  \centering
  \includegraphics[width=\linewidth]{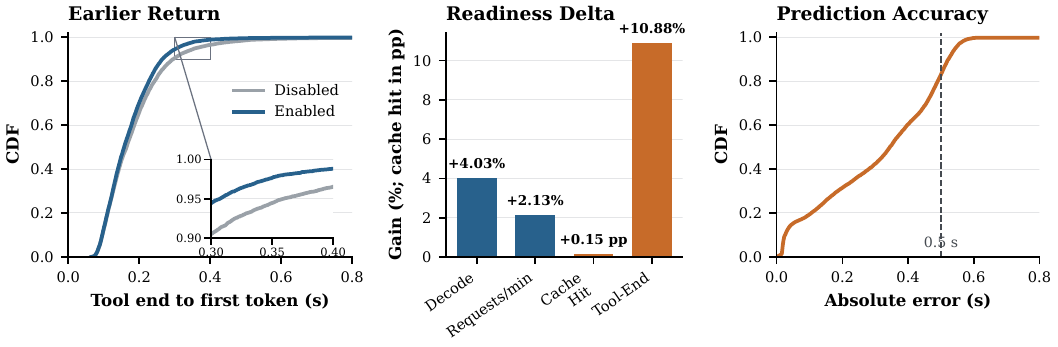}
  \caption{Readiness effectiveness in the matched 32-worker \SWELite{} traces.
  From left to right, the panels show the tool-end-to-first-token CDF, the aggregate
  readiness delta, and absolute error of the structured-history duration
  predictions. Tool-end measures the time from tool completion to the first output token.
  The left inset magnifies 0.3--0.4\,s. The dashed line in the right panel marks
  0.5\,s, half of the 1\,s scheduling period.}
  \label{fig:readiness-effectiveness}
\end{figure}

%% file: sections/experiments/component-ablations.tex
\subsection{Component Ablations}
\label{sec:component-ablations}
Table~\ref{tab:readiness-ablation} compares the complete \Ours{} configuration with variants of its value function, pause ordering, and resumption planner. The complete configuration achieves the highest decode throughput and cache-hit rate, together with the lowest mean \TTFT{}, post-tool delay, and pause wait. Removing cache credit lowers decode throughput from 1417.97 to 1398.35\,tok/s; removing readiness lowers it to 1363.06\,tok/s. Using the pause ordering from \ta{} yields 1355.20\,tok/s, while replacing DP resumption with best-fit-decreasing (BFD) yields 1233.72\,tok/s. Both valuation and decision making contribute to the observed performance.
\input{tables/readiness-ablation.tex}

%% file: tables/readiness-ablation.tex
\begin{table}[t]
\centering
\caption{Component variants on \SWELite{} with DeepSeek-V4 Flash, 2 nodes with $4\times$H100 per node, and 32 workers. Each row uses the indicated component variant of \Ours{}. Decode is in tok/s and hit is prefix-cache reuse (\%). Mean and P90 timing values are reported separately in seconds. The best results are in \textbf{bold} and the second-best results are \underline{underlined}.}
\label{tab:readiness-ablation}
\scriptsize
\setlength{\tabcolsep}{3pt}
\resizebox{\linewidth}{!}{%
\begin{tabular}{@{}lrr*{6}{r}@{}}
\toprule
Configuration & Decode $\uparrow$ & Hit (\%) $\uparrow$ & \multicolumn{2}{c}{\TTFT{} $\downarrow$} & \multicolumn{2}{c}{Tool-end$\to$first $\downarrow$} & \multicolumn{2}{c}{Pause $\downarrow$}\\
\cmidrule(lr){4-5} \cmidrule(lr){6-7} \cmidrule(lr){8-9}
 & & & Mean & P90 & Mean & P90 & Mean & P90\\
\midrule
\textbf{\Ours{} (Ours)} & \textbf{1417.97} & \textbf{96.31} & \textbf{0.154} & \textbf{0.241} & \textbf{0.203} & \textbf{0.299} & \textbf{0.972} & \underline{1.051}\\
\quad \textit{w/o} Cache Value & \underline{1398.35} & \underline{96.24} & \underline{0.157} & \underline{0.246} & \underline{0.209} & \underline{0.309} & 1.028 & \textbf{1.047}\\
\quad \textit{w/o} Readiness & 1363.06 & 96.16 & 0.161 & 0.253 & 0.227 & 0.358 & 1.067 & 1.079\\
\quad \textit{w/o} Cache Value \& Readiness & 1373.51 & \underline{96.24} & 0.182 & 0.278 & 0.213 & 0.317 & \underline{1.025} & 1.059\\
\quad \textit{w/o} Session Evaluation & 1355.20 & 96.20 & 0.162 & 0.257 & 0.226 & 0.350 & \underline{1.025} & 1.061\\
\quad \textit{w/o} Decision Making & 1233.72 & 96.05 & 0.195 & 0.300 & 0.235 & 0.354 & 1.081 & 1.119\\
\bottomrule
\end{tabular}}
\end{table}

%% file: sections/related-work.tex
\section{Related Work}
\label{sec:related}
PagedAttention and RadixAttention improve engine-level KV management, while Preble, Mooncake, and Dynamo exploit distributed cache locality~\citep{vllm,sglang,preble,mooncake,dynamo}. SMetric combines session affinity with load balancing~\citep{smetric}, and production schedulers support extensible scoring~\citep{llmd}. Autellix prioritizes programs using service history, while ThunderAgent introduces the lifecycle and waiting pool on which \Ours{} builds~\citep{autellix,thunderagent}. InferCept and Continuum manage KV retention across tool calls, with Continuum using historical tool-duration distributions~\citep{infercept,continuum}. \Ours{} brings cache opportunity and tool readiness into a joint scheduler-level admission and placement objective, complementing engine-level KV management. Parrot and Helium exploit workflow dependencies, while KVFlow and CacheScout guide residency or prefetching using execution structure~\citep{parrot,helium,kvflow,cachescout}. TOPAS jointly considers prefix residency and admission through bounded search and greedy packing~\citep{topas}. Conveyor, Sutradhara, and PASTE reduce exposed tool latency through partial execution, overlap, or speculation~\citep{conveyor,sutradhara,paste}. \Ours{} converts tool timing into near-term model-service opportunities that inform joint admission and placement. AgentSysBench and MARS motivate accounting for tool delays and deployment boundaries~\citep{agentsysbench,mars}. \Ours{} couples an agentic value interface with joint placement and asynchronous revalidation, using the insights from classical multiple-knapsack and generalized-assignment machinery~\citep{knapsack,shmoystardos}. It becomes popular to formulate some optimization problems in LLM and agentic serving and solve online, from edge scenarios~\citep{li2026moe} to large-scale infrastructure~\citep{bu2026tackling, chen2026online}.

%% file: sections/conclusion.tex
\section{Conclusion}
Tool execution makes agentic inference demand intermittent, while accumulated context makes the choice of node consequential. \Ours{} brings these two properties into a common scheduling objective through cache credit, tool-return readiness, and token footprint. Joint dynamic-programming placement maximizes this objective under snapshot capacity constraints, and asynchronous revalidation connects the plan to live execution. Across the evaluated DeepSeek-V4 Flash and Qwen3-32B settings, this design delivers improvements of up to 26.8\% in latency, 36.3\% in mean \TTFT{}, and 29.5\% in decode throughput. The resulting architecture gives agentic serving an extensible way to translate workload knowledge into coordinated admission and placement.

%% file: sections/prediction-appendix.tex
\section{Historical Tool-Duration Estimator}
\label{app:prediction}
The history-based predictor in Section~\ref{sec:tool-readiness} uses a lightweight summary of completed calls. A finite-sample margin augments the historical mean, allowing the estimate to account for observed variability and the amount of available history.

For calls without direct timing evidence, we maintain statistics of completed durations keyed by the structured tool name $f$. Let $n_f$, $\bar d_f$, $s_f^2$, and $R_f$ denote the sample count, mean, sample variance, and observed duration range. We use the following conservative upper estimate as a duration heuristic~\citep{bernstein}:
\begin{equation}
\label{eq:history-estimator}
\widehat D_f=
\begin{cases}
\bar d_f, & n_f=1,\\[2pt]
\displaystyle\bar d_f+
\sqrt{\frac{2s_f^2\log(3/\delta)}{n_f}}
+\frac{3R_f\log(3/\delta)}{n_f}, & n_f\geq2,
\end{cases}
\end{equation}
with $\delta=0.2$. We use global completed-call statistics when tool-specific history is empty. Subtracting elapsed time from $\widehat D_f$ gives the remaining-duration estimate. The finite-sample margin is a scheduling heuristic; it does not provide calibrated coverage for an individual call. Continuum also exploits tool-duration history to choose a KV-retention TTL~\citep{continuum}. Our estimate converts this historical timing evidence into a signal for scheduler admission.

If both tool-specific and global histories are empty, the implementation uses a five-second duration default. The remaining-time estimate subtracts elapsed time and clips at zero, as in equation~(\ref{eq:history-duration}). Exposed timing hints take precedence as described in the main text.

%% file: sections/method-appendix.tex
\section{Extended Method Details}
\label{app:method-details}

\subsection{Joint selection and placement algorithm}
\input{figures/ours-algorithm.tex}
At every scheduling time, $w_i$ is updated with the memory demand of the total token count minus the shared token capacity for session $i$, and $C_b$ is updated with the remaining capacity of node $b$ from the metric observation. Algorithm~\ref{alg:ours} summarizes \Ours{}, from \ValueAssign{} to joint scheduling and live-state revalidation.

\subsection{Asynchronous execution}
\label{sec:async-execution}
The scheduler solves a frozen snapshot in an isolated worker. Before applying the plan, it checks that selected \Sessions{} remain eligible and match their recorded state signatures, then applies the longest prefix permitted by current capacities. Changed selected-\Session{} signatures cause the plan to be discarded. Revalidation preserves feasibility under current admission constraints. We quantify the effect of stale values, discarded plans, and partial application through a loss bound relative to the fresh-snapshot optimum, and measure these execution outcomes alongside solve time and plan age.
\input{sections/async-bound.tex}


%% file: figures/ours-algorithm.tex
\begin{figure}[H]
\refstepcounter{algorithm}
\label{alg:ours}
\hrule
\smallskip
\noindent\textbf{Algorithm \thealgorithm: \Ours{} joint selection and placement}
\smallskip
\hrule
\small
\begin{tabbing}
\hspace{1.2em}\=\hspace{1.2em}\=\hspace{1.2em}\=\hspace{1.2em}\=\kill
\textbf{Input:} frozen candidates $P$, nodes $B$, capacities $C$, eligibility $E$, quantum $\Delta$\\
\textbf{Output:} a revalidated set of \Session{}--node assignments\\
Compute $w_i$ and $\boldsymbol v_{i,b}=(k_{i,b},r_i,t_i)$ through \ValueAssign{}\\
$\bar w_i\gets\lceil w_i/\Delta\rceil$;\quad $\bar C_b\gets\lfloor C_b/\Delta\rfloor$\\
Embed each value to obtain $\boldsymbol u_{i,b}$\\
Initialize $D_0[\boldsymbol 0]\gets\boldsymbol 0$; other states are unreachable\\
\textbf{for} candidate $j=1,\ldots,|P|$ \textbf{do}\\
\> $D_j\gets D_{j-1}$, recording skip predecessors\\
\> \textbf{for each} reachable state $\boldsymbol s$ in $D_{j-1}$ \textbf{do}\\
\>\> \textbf{for each} eligible node $b$ with $s_b+\bar w_j\leq\bar C_b$ \textbf{do}\\
\>\>\> $\boldsymbol s'\gets\boldsymbol s+\bar w_j\boldsymbol e_b$\\
\>\>\> $\boldsymbol z\gets D_{j-1}[\boldsymbol s]+\boldsymbol u_{j,b}$\\
\>\>\> \textbf{if} $\boldsymbol s'$ is new or $\boldsymbol z$ improves $D_j[\boldsymbol s']$ \textbf{then}\\
\>\>\>\> Store $\boldsymbol z$ and predecessor $(\boldsymbol s,j,b)$\\
\> If the budget is exceeded, obtain a BFD plan and skip to revalidation\\
Otherwise choose the best final state and backtrack to recover the plan\\
\textbf{Revalidation:}\\
Discard the plan if a selected \Session{}'s live signature or eligibility changed\\
\textbf{return} the longest plan prefix feasible under current admission capacities
\end{tabbing}
\hrule
\end{figure}

%% file: sections/async-bound.tex
\paragraph{Conditional loss bound.}
Let $s$ and $t$ be planning and application times, $x_s$ the exact snapshot plan,
and $y_t$ the revalidated plan (empty if discarded). Write $\mathcal X_s^\Delta$
and $\mathcal X_t^\Delta$ for the quantized feasible sets. For analysis, represent
the lexicographic objective by one fixed scalar score $U_\tau$ that preserves its
order on the finite plans and values under comparison; the implementation still
compares vectors. Define
\begin{align}
\epsilon_{s,t}
&=\max_{x\in\mathcal X_s^\Delta}|U_t(x)-U_s(x)|,\nonumber\\
A_{s,t}
&=\max_{x\in\mathcal X_t^\Delta}U_t(x)
-\max_{x\in\mathcal X_s^\Delta\cap\mathcal X_t^\Delta}U_t(x),\nonumber\\
T_{s,t}&=[U_t(x_s)-U_t(y_t)]_+.
\label{eq:async-loss-terms}
\end{align}
These terms capture value drift, opportunities introduced by feasibility changes,
and value lost during execution. The empty plan ensures a nonempty intersection.
Then
\begin{equation}
\label{eq:async-bound}
\max_{x\in\mathcal X_t^\Delta}U_t(x)-U_t(y_t)
\leq A_{s,t}+2\epsilon_{s,t}+T_{s,t}.
\end{equation}
The proof in Appendix~\ref{app:async-proof} inserts the best common feasible plan
and applies snapshot optimality plus two value-drift terms. If every plan has
at most $K$ assignments and each scalar assignment value changes by at most
$\eta_{s,t}$, then $\epsilon_{s,t}\leq K\eta_{s,t}$. With an additional drift bound
$\eta_{s,t}\leq L(t-s)$, the value-drift term becomes $2KL(t-s)$; readiness jumps require separate allowances. An unquantized comparator adds the
quantization gap, and a fallback adds its snapshot suboptimality gap.

The bound connects planner freshness to retained scheduling value: short plan
age limits value drift under the stated assumptions, while revalidation losses
appear explicitly in $T_{s,t}$. Its comparator is the fresh placement optimum
on the same realized state; Appendix~\ref{app:async-proof} details this scope.

%% file: sections/async-bound-proof.tex
\section{Asynchronous Snapshot Loss: Proof and Scope}
\label{app:async-proof}

\paragraph{Loss relative to a fresh snapshot.}
We compare the applied plan with the optimal placement at application time on the same observed system state. Let $s$ be the planning time and $t$ the application time. Write $\mathcal X_s^\Delta$ and $\mathcal X_t^\Delta$ for the discretized feasible sets, embedding both in the union of candidate--node assignments. The empty plan belongs to both sets. Let $x_s$ be the exact snapshot solution and $y_t\in\mathcal X_t$ the applied plan, feasible under the current admission contract; a discarded plan has $y_t=0$.

To express loss numerically, let $U_\tau(x)$ be a fixed positive weighted sum of the objective coordinates in equation~(\ref{eq:objective}), with weights that preserve their lexicographic order on the finite collection of plans and objective values considered at $s$ and $t$. Such weights exist for finite sets, with a scale determined by the coordinate ranges and gaps. The scalar score serves the analysis; the implementation compares vectors directly. In particular, $x_s$ maximizes $U_s$ over $\mathcal X_s^\Delta$. Extend the additive objective to all assignments in the union, including plans that have become infeasible.

Define the value variation, loss from changes in feasibility, and execution loss as
\begin{align}
\epsilon_{s,t}
&=\max_{x\in\mathcal X_s^\Delta}|U_t(x)-U_s(x)|,\nonumber\\
A_{s,t}
&=\max_{x\in\mathcal X_t^\Delta}U_t(x)
 -\max_{x\in\mathcal X_s^\Delta\cap\mathcal X_t^\Delta}U_t(x),\nonumber\\
T_{s,t}&=[U_t(x_s)-U_t(y_t)]_+.
\label{eq:async-loss-terms-proof}
\end{align}
Here $[z]_+=\max(z,0)$. New candidates, new eligibility, and additional capacity can contribute to $A_{s,t}$; truncation or discard contributes to $T_{s,t}$. Changes in feature values contribute to $\epsilon_{s,t}$.

\paragraph{Conditional bound.}
For every exact snapshot solve and feasible applied plan,
\begin{equation}
\label{eq:async-bound-proof}
\max_{x\in\mathcal X_t^\Delta}U_t(x)-U_t(y_t)
\leq A_{s,t}+2\epsilon_{s,t}+T_{s,t}.
\end{equation}
To prove this, choose $z$ maximizing $U_t$ over the intersection of the two feasible sets. Snapshot optimality gives $U_s(z)\leq U_s(x_s)$, and hence
\begin{align*}
U_t(z)-U_t(y_t)
&\leq [U_t(z)-U_s(z)]
 +[U_s(x_s)-U_t(x_s)]\\
&\quad +[U_t(x_s)-U_t(y_t)]
\leq 2\epsilon_{s,t}+T_{s,t}.
\end{align*}
Adding $A_{s,t}$ proves equation~(\ref{eq:async-bound-proof}). If the comparator uses the original, undiscretized capacities, add the nonnegative quantization gap
$Q_t=\max_{x\in\mathcal X_t}U_t(x)-\max_{x\in\mathcal X_t^\Delta}U_t(x)$
to its right-hand side.

The bound isolates conditions under which asynchronous planning retains value. If every plan contains at most $K$ assignments and each scalar assignment value changes by at most $\eta_{s,t}$, then $\epsilon_{s,t}\leq K\eta_{s,t}$. A further assumption $\eta_{s,t}\leq L(t-s)$ yields a variation term $2KL(t-s)$; binary readiness require explicit jump allowances and do not satisfy this assumption automatically. Summing equation~(\ref{eq:async-bound-proof}) over scheduling ticks bounds cumulative loss against fresh-snapshot optima on the realized states, using one common scalarization. It does not bound end-to-end latency or regret against a globally optimal online policy, whose earlier decisions can produce different future states. A uniform numerical bound requires corresponding bounds on state variation, feasibility changes, and discarded value. A fallback solve additionally contributes its snapshot suboptimality gap.

%% file: sections/glm-traces-appendix.tex
\section{Additional GLM-4.7 Serving Traces}
\label{app:glm-traces}
Figures~\ref{fig:glm-traces-w96} and~\ref{fig:glm-traces-w128} extend Figure~\ref{fig:glm-traces} to 96 and 128 workers on the 300-task \SWELite{} workload with GLM-4.7 TP8. Each trace starts at its run's first client request and shows the observed number of running backend requests. In both settings, \Ours{} reaches the end of the observed workload window earlier than the displayed baselines. The curves describe natural-live runs with differing generated trajectories.
\input{figures/glm-traces-appendix.tex}

%% file: figures/glm-traces-appendix.tex
\begin{figure}[H]
\centering
\includegraphics[width=\linewidth]{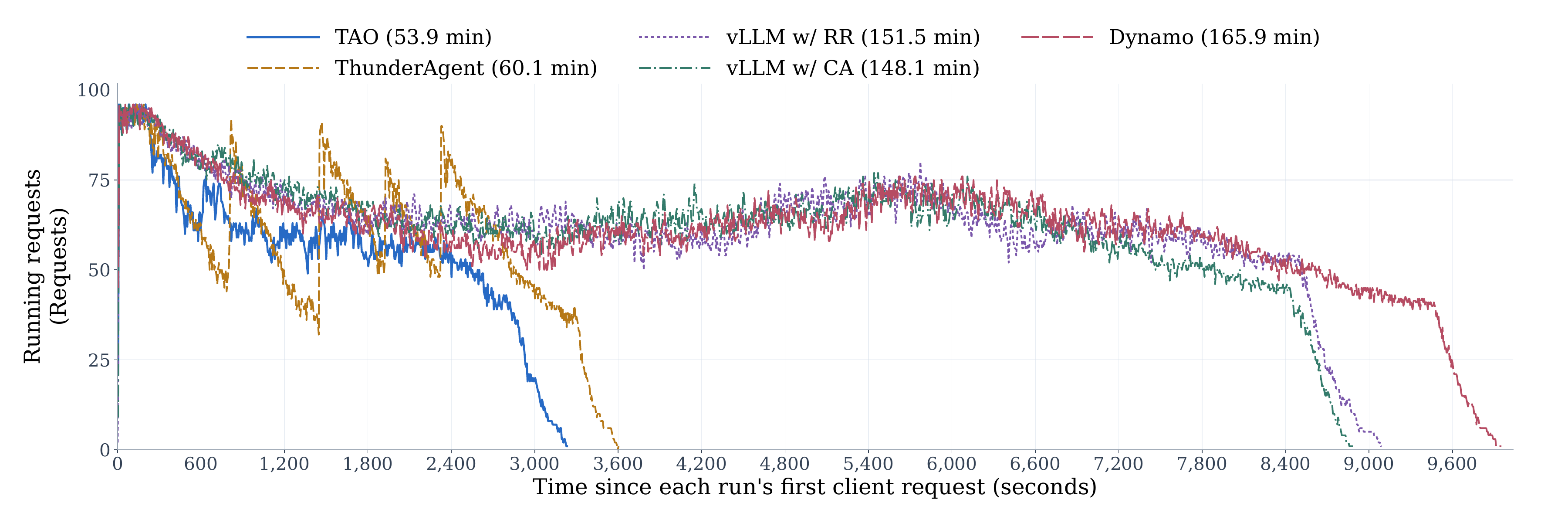}
\caption{GLM-4.7 running-request traces on \SWELite{} at 96 workers (TP8), aligned to each run's first client request.}
\label{fig:glm-traces-w96}
\end{figure}

\begin{figure}[H]
\centering
\includegraphics[width=\linewidth]{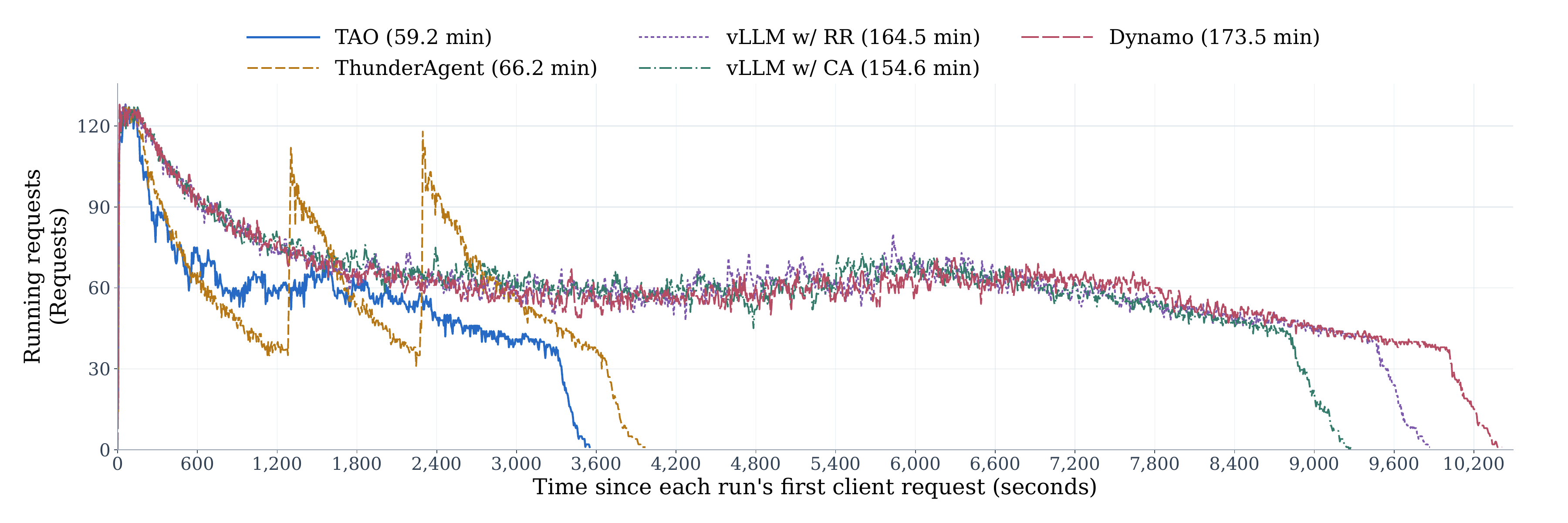}
\caption{GLM-4.7 running-request traces on \SWELite{} at 128 workers (TP8), aligned to each run's first client request.}
\label{fig:glm-traces-w128}
\end{figure}

%% file: sections/tp8-appendix.tex
\subsection{Fallback Compatibility}
\label{app:tp8-miniswe}
Table~\ref{tab:tp8-miniswe-appendix} evaluates five schedulers on a single
TP8 node across 4--64 workers. Each run serves the 300-task \SWELite{}
workload through MiniSWE. This experiment checks that \Ours{} remains
competitive when concurrency is low and scheduling has less opportunity to
improve performance. At 4--8 workers, its mean task latency stays within
2.3\% of the best baseline, it retains over 92\% of the best baseline's
decode throughput, and its mean \TTFT{} is only 15--20\,ms above the best
baseline. Thus, \Ours{} incurs little task-latency degradation at low
concurrency while maintaining competitive throughput. Together with its
leading performance in the main evaluation
(Section~\ref{sec:main-results}), this fallback behavior supports choosing
\Ours{} as the default scheduler across the evaluated operating regimes.
\input{tables/tp8-miniswe-appendix.tex}

%% file: tables/tp8-miniswe-appendix.tex
\begin{table}[t]
\centering
\caption{Low-concurrency fallback check for DeepSeek-V4 Flash on \SWELite{}, one TP8 node ($8\times$H100), with a 4--64-worker sweep. Each run serves 300 tasks. Latency is the mean task E2E time, TTFT is the mean client time to first token, and decode is aggregate throughput. Boldface and underlining indicate the best and second-best distinct values.}
\label{tab:tp8-miniswe-appendix}
\scriptsize
\setlength{\tabcolsep}{5pt}
\begin{tabular}{c l r r r}
\toprule
Workers & Method & E2E latency (s) & TTFT (s) & Decode (tok/s)\\
\midrule
4  & \vrr{} &  {87.911} & 0.126 &  {317.855} \\
4  & \vca{} &  {88.390} & 0.125 &  {318.357} \\
4  & \dy{}  & 92.549 &  {0.102} & 307.382 \\
4  & \ta{}  & 90.099 & 0.120 & 312.368 \\
4  & \Ours{} (Ours) & 89.479 &  {0.117} & 315.629 \\
\midrule
8  & \vrr{} & 102.868 & 0.139 & 528.074 \\
8  & \vca{} &  {101.698} & 0.142 &  {555.818} \\
8  & \dy{}  & 103.813 &  {0.112} &  {565.893} \\
8  & \ta{}  &  {100.610} &  {0.130} & 527.068 \\
8  & \Ours{} (Ours)& 102.913 & 0.132 & 522.838 \\
\midrule
16 & \vrr{} &  {121.158} & 0.160 & 831.120 \\
16 & \vca{} & 126.907 & 0.176 & 861.089 \\
16 & \dy{}  &  {126.670} &  {0.130} &  {884.575} \\
16 & \ta{}  & 147.709 & 0.789 & 722.445 \\
16 & \Ours{} (Ours)& 126.702 &  {0.153} &  {861.198} \\
\midrule
32 & \vrr{} &  {160.981} & 0.211 & 1085.654 \\
32 & \vca{} &  {166.657} & 0.218 &  {1232.259} \\
32 & \dy{}  & 172.228 &  {0.171} & 1147.924 \\
32 & \ta{}  & 272.845 & 3.919 & 766.776 \\
32 & \Ours{} (Ours)& 171.389 &  {0.195} &  {1240.768} \\
\midrule
64 & \vrr{} &  {225.967} &  {0.278} &  {1788.345} \\
64 & \vca{} &  {225.513} & 0.279 & 1658.347 \\
64 & \dy{}  & 229.670 &  {0.231} &  {1704.425} \\
64 & \ta{}  & 475.033 & 9.332 & 794.991 \\
64 & \Ours{} (Ours)& 226.268 & 0.423 & 1667.524 \\
\bottomrule
\end{tabular}
\end{table}

%% file: sections/tokencake-appendix.tex
\subsection{KV-Cache Management Compatibility}
\label{app:tokencake}

\TokenCake{} uses workflow dependencies and agent priorities to guide its Space Scheduler, and tool-stall events and timing prediction to guide offload and upload in its Time Scheduler~\citep{tokencake}. Agentix prioritizes and preempts LLM calls using program-level service history to reduce request- and program-level blocking~\citep{agentix}. It operates at the request-scheduling layer, so integrating it with \Ours{} would require coordinating the two admission policies. KVFlow represents a workflow as an Agent Step Graph and uses steps-to-execution to prioritize cache eviction and prefetching~\citep{kvflow}. CacheScout learns execution transitions online to guide cache management without requiring a predefined graph~\citep{cachescout}. These mechanisms provide information about future cache use. In our formulation, such information can inform the program--node values supplied to \ValueAssign{}, while engine-local cache policies can operate beneath \Ours{}'s admission and placement decisions. This is an architectural extension for KVFlow and CacheScout, while the implemented composition evaluated here uses \TokenCake{}.

With \TokenCake{} enabled, the combined system completes validated 300-task \SWELite{} runs on DeepSeek-V4 Flash TP4$\times$2 at 16 and 32 workers. Relative to the same engine behind consistent-hash routing, the combination increases decode throughput from 785.159 to 886.772 tok/s and from 1292.584 to 1347.373 tok/s, respectively (12.9\% and 4.2\%). These runs establish operational compatibility.

\input{tables/tokencake-compatibility.tex}
Table~\ref{tab:tokencake-compatibility} compares three frontends over the \TokenCake{} engine. At 16 and 32 workers, \Ours{} lowers mean task latency relative to consistent hashing by 3.1\% and 2.0\%, while increasing decode throughput by 12.9\% and 4.2\%. At 32 workers, client TTFT increases slightly from 0.220 to 0.223\,s. Both combined runs record balanced request routing across the two replicas (5816/5393 and 5328/5683 requests) and tool-prediction updates (22085 and 21776). Neither records a program pause, resume, or migration, and their final offload series are absent. The evidence therefore establishes lifecycle and serving compatibility under these loads; attribution to DP selection or overlapping offload would require runs that activate those mechanisms. Separate transfer qualifications verify actual GPU--CPU movement.

%% file: tables/tokencake-compatibility.tex
\begin{table}[t]
\centering
\caption{Frontend composition over \TokenCake{} on DeepSeek-V4 Flash TP4$\times$2, \SWELite{} (300 tasks), with 100\,GiB total CPU KV capacity. Consistent-hash rows are retained validated runs; cache-aware and \Ours{} rows are from the later campaign. Bold and underline rank the TokenCake configurations within each worker group. Light-blue rows reproduce the original TAO results from Table~\ref{tab:main-results} for reference and are excluded from ranking and relative-improvement calculations.}
\label{tab:tokencake-compatibility}
\footnotesize
\setlength{\tabcolsep}{5pt}
\begin{tabular}{@{}rlrrr@{}}
\toprule
Workers & Frontend & Latency $\downarrow$ (s) & TTFT $\downarrow$ (s) & Decode $\uparrow$ (tok/s)\\
\midrule
16 & \TokenCake{} & \underline{126.891} & \underline{0.178} & \underline{785.159}\\
 & Cache-aware w/ \TokenCake{} & 137.907 & 0.268 & 760.166\\
 & \Ours{} w/ \TokenCake{} & \textbf{122.940} & \textbf{0.177} & \textbf{886.772}\\
\rowcolor{gray!15}
 & Rel. Imprv. (\%) & +3.1 & +0.7 & +12.9\\
\rowcolor{blue!8}
 & \Ours{} (Ours) & 113.264 & 0.135 & 934.967\\
\midrule
32 & \TokenCake{} & \underline{162.905} & \textbf{0.220} & \underline{1292.584}\\
 & Cache-aware w/ \TokenCake{} & 284.667 & 0.715 & 778.611\\
 & \Ours{} w/ \TokenCake{} & \textbf{159.581} & \underline{0.223} & \textbf{1347.373}\\
\rowcolor{gray!15}
 & Rel. Imprv. (\%) & +2.0 & -1.5 & +4.2\\
\rowcolor{blue!8}
 & \Ours{} (Ours) & 142.332 & 0.153 & 1477.463\\
\midrule
64 & \TokenCake{} & \underline{509.018} & \underline{1.590} & \underline{790.491}\\
 & Cache-aware w/ \TokenCake{} & 695.986 & 3.022 & 613.004\\
 & \Ours{} w/ \TokenCake{} & \textbf{274.892} & \textbf{0.593} & \textbf{1414.547}\\
\rowcolor{gray!15}
 & Rel. Imprv. (\%) & +46.0 & +62.7 & +78.9\\
\rowcolor{blue!8}
 & \Ours{} (Ours) & 185.410 & 0.224 & 2021.523\\
\bottomrule
\end{tabular}
\end{table}

%% file: sections/experiments/inferencex-replay.tex
\section{InferenceX Replay}
\label{sec:inferencex-replay}
\Ours{} also achieves the highest decode throughput at every tested concurrency level on InferenceX, a fixed trace corpus served by DeepSeek on 2 nodes with $4\times$H100 per node. The replay uses 393 AgentX traces with a 1M-token context limit and 1,800-second profiling windows~\citep{agentx,inferencex}; workers count the concurrent session trees. Table~\ref{tab:inferencex-replay} reports client-visible \TTFT{} and decode throughput, together with seconds per request computed as $60/R$, where $R$ is native request throughput in requests per minute. This last metric is the reciprocal of aggregate request throughput within the profiling window.
\input{tables/inferencex-replay.tex}
Decode throughput improves on the best competing method by 4.2\%, 1.8\%, and 10.1\% at 16, 32, and 64 workers, respectively. \Ours{} also achieves the lowest \TTFT{} at 16 workers; cache-aware scheduling has lower \TTFT{} at 32 and 64 workers. Figure~\ref{fig:inferencex-cache-hit-tp4} shows the accompanying reuse advantage: \Ours{} maintains the highest prefix-cache hit rate throughout, even as hit rates decrease with concurrency for all four methods.
\input{figures/inferencex-tp4-cache-hit/hit-rate.tex}

%% file: tables/inferencex-replay.tex
\begin{table}[t]
\centering
\caption{InferenceX replay on 2 nodes with $4\times$H100 per node. Latency is reported in seconds per request (s/req). TTFT is recovered client time (s), and decode is node-window decode TPS. These replay measurements use a fixed trace corpus and are reported separately from natural-live task results.}
\label{tab:inferencex-replay}
\small
\setlength{\tabcolsep}{3pt}
\resizebox{\linewidth}{!}{%
\begin{tabular}{l*{9}{r}}
\toprule
Method & \multicolumn{3}{c}{16 workers} & \multicolumn{3}{c}{32 workers} & \multicolumn{3}{c}{64 workers}\\
\cmidrule(lr){2-4}\cmidrule(lr){5-7}\cmidrule(l){8-10}
 & Latency (s/req) & TTFT (s) & Decode (tok/s) & Latency (s/req) & TTFT (s) & Decode (tok/s) & Latency (s/req) & TTFT (s) & Decode (tok/s)\\
\midrule
\dy{} & \underline{5.545} & 10.844 & \underline{133.158} & \underline{3.234} & 26.542 & 204.257 & 5.154 & 163.674 & 71.085\\
\vrr{} & 7.121 & 10.861 & 107.478 & 5.069 & 23.058 & 127.991 & 4.589 & 45.852 & 87.050\\
\vca{} & 5.916 & \underline{7.421} & 115.947 & \textbf{3.103} & \textbf{9.006} & \underline{211.262} & \underline{3.825} & \textbf{25.145} & \underline{109.413}\\
\Ours{} (Ours) & \textbf{5.382} & \textbf{4.756} & \textbf{138.701} & \underline{3.234} & \underline{10.302} & \textbf{215.075} & \textbf{3.532} & \underline{27.190} & \textbf{120.469}\\
\rowcolor{gray!15}
\textit{Rel. Imprv. (\%)} & +2.9 & +35.9 & +4.2 & -4.2 & -14.4 & +1.8 & +7.7 & -8.1 & +10.1\\
\bottomrule
\end{tabular}
}
\end{table}

%% file: figures/inferencex-tp4-cache-hit/hit-rate.tex
\begin{figure}[H]
  \centering
  \includegraphics[width=0.52\linewidth]{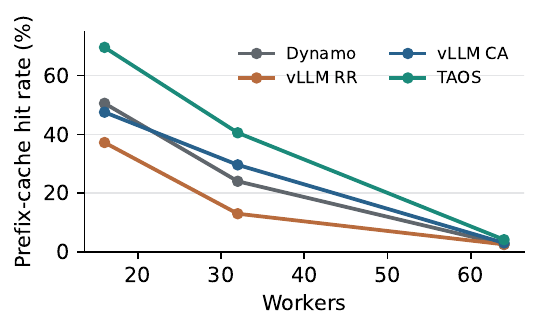}
  \caption{InferenceX prefix-cache hit rate on 2 nodes with $4\times$H100 per node. \Ours{} maintains higher cache reuse across the tested worker counts.}
  \label{fig:inferencex-cache-hit-tp4}
\end{figure}

%% file: sections/tool-delay-appendix.tex
\section{Sensitivity to longer tool-return delays}
\label{app:tool-delay}
We examine how longer tool-return delays affect scheduling using a replay of \SWELite{} sessions recorded with MiniSWE. The replay fixes the source
prompts, 6,995 model-request records, and 6,942 tool records, while model
outputs are generated live. Each session receives a fixed additional
tool-return delay drawn from 1--5\,s; multipliers of 1, 2, and 4 produce
additional delays of 1--5, 2--10, and 4--20\,s, respectively.
DeepSeek-V4 Flash runs on TP4$\times$2 with 32 workers, a 33,024-token
context limit, and a 512-token per-session reserve.

Table~\ref{tab:tool-delay-scaling} compares three configurations: the historical
\Ours{} w/o readiness reference, \Ours{} with structured-history tool-duration
predictions, and Oracle with the replay's known effective tool
duration. The historical reference also disables cached-token decay and
shared-token refresh, which both current configurations enable; comparisons
with that reference therefore measure the combined configuration change.
Acting-token decay remains disabled in both current configurations.
All six current cells pass the replay validator, covering
all 6,942 tool records. Each cell is a single exploratory run with
batch-invariant execution disabled and live model generation.

\paragraph{\Ours{} sustains gains as tool delays grow.}
Relative to the historical configuration, \Ours{} reduces makespan by
20.72\%, 16.42\%, and 19.28\% and mean session latency by 21.57\%,
20.57\%, and 17.91\% at scales 1, 2, and 4, respectively. These gains
span additional tool-return delays from 1--5\,s to 4--20\,s.

\paragraph{Longer delays make readiness relevant to more placement decisions.}
For \Ours{}, removing readiness from the same captured state changes
11/668 (1.65\%), 53/1,345 (3.94\%), and 200/2,521 (7.93\%) of the
exact DP plan comparisons at multipliers 1, 2, and 4. The rate increases by
6.29 percentage points from scale 1 to scale 4, indicating that tool-return
timing affects a growing fraction of placement decisions as delays lengthen.
These same-state comparisons isolate readiness's effect on individual
placement decisions within the \Ours{} runs.

\paragraph{History-based predictions approach oracle runtime performance.}
Providing the known replay duration reduces mean absolute prediction error
by 99.53--99.69\%, to 0.025--0.030\,s. At scale 2, it also reduces
makespan by 2.24\% and mean session latency by 0.71\% relative to
\Ours{}. Across all three scales, the signed makespan improvements range
from $-0.10$\% to +2.24\%, and mean session-latency improvements range
from $-0.36$\% to +0.71\%. Thus, the history-based predictor achieves
runtime performance close to the accurate-timing variant in this sweep,
even as longer delays increase the frequency of readiness-sensitive decisions.
\input{tables/tool-delay-scaling.tex}

%% file: tables/tool-delay-scaling.tex
\begin{table}[t]
\centering
\caption{Tool-return delay scaling on \SWELite{} replay sessions. Scales 1, 2, and 4 add 1--5, 2--10, and 4--20\,s per tool return, respectively. Oracle uses accurate timing supplied by the known replay tool duration. MAE is logged duration-prediction error, including predictions unused by the no-readiness arm. Decision change is the percentage of exact snapshot comparisons whose plan changes when readiness is removed. Rel. Imprv. compares \Ours{} with \Ours{} w/o readiness: positive percentages denote makespan and mean E2E reductions; the last column gives the decision-change difference in percentage points (pp). Oracle results are a timing-information reference.}
\label{tab:tool-delay-scaling}
\scriptsize
\setlength{\tabcolsep}{4pt}
\begin{tabular}{@{}r l r r r r@{}}
\toprule
Scale & Prediction arm & Makespan (s) & Mean E2E (s) & MAE (s) & Decision change (\%)\\
\midrule
1 & \Ours{} w/o readiness & 1,919.906 & 229.714 & 6.531 & 0.000 \\
1 & \Ours{} & 1,522.076 & 180.167 & 9.490 & 1.647 \\
1 & \textit{Oracle} & 1,523.532 & 180.030 & 0.030 & 4.167 \\
 & Rel. Imprv. & $+20.72\%$ & $+21.57\%$ & -- & $+1.65$ pp \\
\midrule
2 & \Ours{} w/o readiness & 2,816.496 & 352.502 & 2.593 & 0.000 \\
2 & \Ours{} & 2,354.050 & 279.986 & 6.315 & 3.941 \\
2 & \textit{Oracle} & 2,301.282 & 278.011 & 0.030 & 4.059 \\
 & Rel. Imprv. & $+16.42\%$ & $+20.57\%$ & -- & $+3.94$ pp \\
\midrule
4 & \Ours{} w/o readiness & 5,090.251 & 599.155 & 4.512 & 0.000 \\
4 & \Ours{} & 4,108.865 & 491.829 & 5.837 & 7.933 \\
4 & \textit{Oracle} & 4,110.209 & 493.595 & 0.025 & 7.398 \\
 & Rel. Imprv. & $+19.28\%$ & $+17.91\%$ & -- & $+7.93$ pp \\
\bottomrule
\end{tabular}
\end{table}

%% file: sections/placement-case-studies.tex
\section{Joint Placement Case Studies}
\label{app:placement-cases}
Reconstructed snapshots from the 32-worker DeepSeek replay show how joint placement preserves more cache value and admits more sessions than a short-first heuristic (Section~\ref{sec:improvement-breakdown}). The figures compare scheduler-level assignment decisions and their cache credit; vLLM manages the physical KV cache. The short-first heuristic is evaluated offline on the same snapshots.

A representative cache-credit snapshot is shown in Figure~\ref{fig:dp-cache-safe-advantage}.
DP retains 16,791 tokens of cache credit, compared with 13,624 for the
short-first heuristic, a 23.2\% increase with the same total safe-token
remainder. A separate reconstructed snapshot in Figure~\ref{fig:dp-placement}
shows the selection effect: DP admits seven sessions, including long-context
session~177, while the heuristic admits six and defers that session.
\input{figures/dp-cache-safe-advantage.tex}
\input{figures/dp-placement.tex}

%% file: figures/dp-cache-safe-advantage.tex
\begin{figure}[t]
  \centering
  \includegraphics[width=\linewidth]{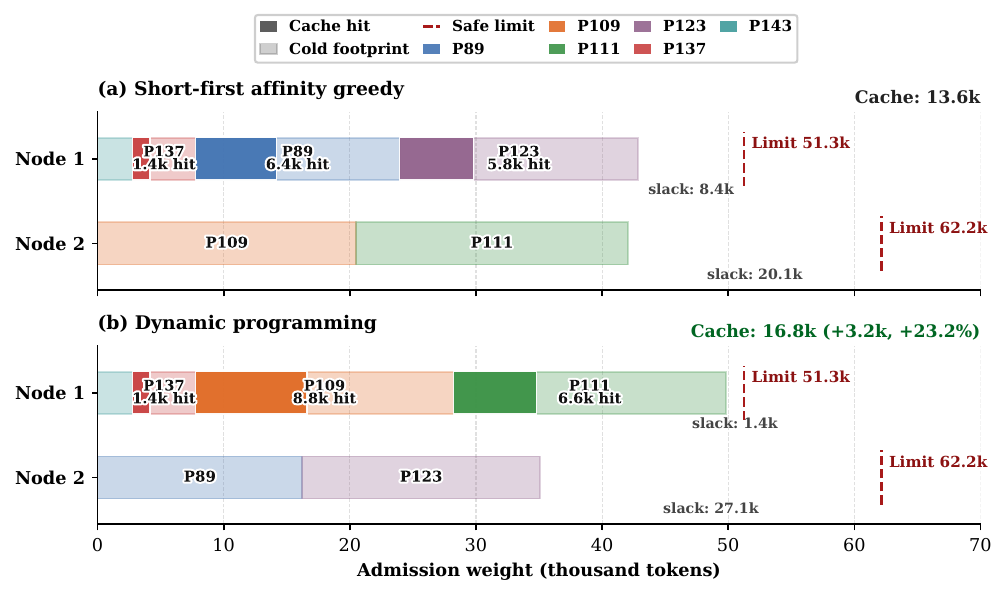}
  \caption{Cache utilization in one placement snapshot. Each program has a distinct hue; the translucent segment is its admission weight and the saturated overlay represents retained cache credit (shown as token labels when space permits). DP retains 16,791 cache-credit tokens versus 13,624 for short-first placement (23.2\% more), with the same total safe-token remainder.}
  \label{fig:dp-cache-safe-advantage}
\end{figure}

%% file: figures/dp-placement.tex
\begin{figure}[t]
  \centering
  \includegraphics[width=\linewidth]{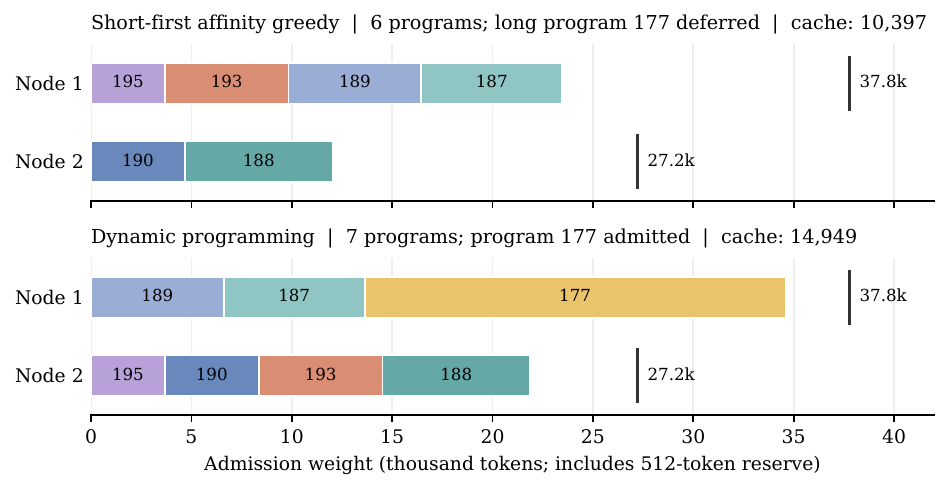}
  \caption{A reconstructed replay snapshot in which DP admits seven sessions while the short-first affinity-greedy heuristic admits six. Candidate widths represent admission weights, including the 512-token reserve. DP admits the long-context session~177; the heuristic leaves it deferred after filling capacity with shorter programs.}
  \label{fig:dp-placement}
\end{figure}

%% file: sections/overhead-appendix.tex
\section{Scheduler Overhead Analysis}
\label{app:overhead-details}

Across the four-replica Qwen runs at 96--160 workers, planning submissions rise from 98 to 1,681. At 160 workers, completed assignment solving averages 1.59\,ms with a P95 of 3.68\,ms; the two measured parts together account for 4.03\,s (0.112\% of the run). Thus, the recorded scheduling work remains below 0.12\% at four replicas and up to 160 workers.


Table~\ref{tab:overhead-analysis-full} expands the scheduling work and planner outcomes summarized above. State preparation runs synchronously in the scheduler, while the planner worker solves assignments asynchronously. Result delay measures dispatch, solving, and delivery together. The two- and four-replica deployments exercise this design under different models and concurrency levels.

The state budget bounds DP search, with best-fit-decreasing placement handling exhausted searches. Fallback occurs in 5.24\% of submissions at 64 DeepSeek workers and 3.06--8.98\% in the Qwen sweep. Measured work covers state preparation and completed solves; exhausted searches and fallback execution are untimed, so work/run is a partial accounting rather than a measure of CPU utilization or request-latency overhead. Discards and partial applications record how live revalidation handles state changes before application. The 16-worker DeepSeek run submits one plan, so its timing entry describes that single solve.

\input{tables/overhead-analysis-appendix.tex}

%% file: tables/overhead-analysis-appendix.tex
\begin{table}[t]
\centering
\caption{Scheduler overhead on \SWELite{} (300 tasks per run). TP4$\times$2 uses two replicas; TP2$\times$4 uses four. State preparation is the time to collect the current candidates and capacities; assignment solving is the completed asynchronous planning time; result delay includes dispatch and delivery. Measured work sums state preparation and completed solving, excluding failed attempts and BFD execution. Work fractions divide by run duration. Outcome entries are count (\%): fallback/discard rates divide by submissions, partial rates by application events (including empty plans).}
\label{tab:overhead-analysis-full}
\small
\setlength{\tabcolsep}{3pt}
\begin{tabular*}{\linewidth}{@{\extracolsep{\fill}}lrrrrrr@{}}
\toprule
 & \multicolumn{3}{c}{DeepSeek-V4 Flash, TP4$\times$2} & \multicolumn{3}{c}{Qwen3-32B, TP2$\times$4} \\
\cmidrule(lr){2-4}\cmidrule(l){5-7}
Workers & 16 & 32 & 64 & 96 & 128 & 160 \\
\midrule
Run duration (s) & 2321.1 & 1514.2 & 1199.0 & 3151.2 & 3077.9 & 3583.0 \\
State preparation total (ms) & 1.37 & 30.58 & 394.95 & 50.67 & 1024.09 & 1588.17 \\
Assignment solving total (ms) & 0.55 & 48.97 & 3163.36 & 52.77 & 1351.51 & 2437.68 \\
Assignment solving mean (ms) & 0.55 & 1.48 & 12.50 & 0.56 & 1.04 & 1.59 \\
Assignment solving p95 (ms) & 0.55 & 2.08 & 88.84 & 0.62 & 2.43 & 3.68 \\
Result delay mean (ms) & 337.49 & 20.83 & 27.86 & 6.07 & 4.23 & 4.89 \\
Result delay p95 (ms) & 337.49 & 22.83 & 105.09 & 5.29 & 9.03 & 9.99 \\
\midrule
State preparation/run (\%) & 0.00006 & 0.00202 & 0.03294 & 0.00161 & 0.03327 & 0.04433 \\
Assignment solving/run (\%) & 0.00002 & 0.00323 & 0.26384 & 0.00167 & 0.04391 & 0.06804 \\
Measured scheduler work/run (\%) & 0.00008 & 0.00525 & 0.29678 & 0.00328 & 0.07718 & 0.11236 \\
\midrule
Plan submissions & 1 & 33 & 267 & 98 & 1395 & 1681 \\
Completed plans & 1 & 33 & 253 & 95 & 1301 & 1530 \\
Plan applications & 1 & 33 & 252 & 95 & 1299 & 1527 \\
Nonempty applications & 1 & 33 & 240 & 89 & 678 & 687 \\
BFD fallback & 0 (0.00) & 0 (0.00) & 14 (5.24) & 3 (3.06) & 94 (6.74) & 151 (8.98) \\
Discarded & 0 (0.00) & 0 (0.00) & 1 (0.37) & 0 (0.00) & 2 (0.14) & 3 (0.18) \\
Partial applications & 0 (0.00) & 0 (0.00) & 2 (0.79) & 0 (0.00) & 0 (0.00) & 0 (0.00) \\
\bottomrule
\end{tabular*}
\end{table}

%% file: sections/related-work-appendix.tex
\section{Extended Related Work}
\label{app:related-work}

\paragraph{Capacity-constrained optimization.}
Dynamic programming and exact multiple-knapsack algorithms provide established tools for selecting items and assigning them to capacity-limited resources~\citep{bellman,knapsack,pisinger}. Node-dependent assignment values place our formulation within the multiple-knapsack/generalized-assignment family. Approximation algorithms address the computational cost of this family~\citep{chekurikhanna,shmoystardos}, while online resource-allocation algorithms analyze decisions under specified arrival models~\citep{lisunye}. \Ours{} uses this classical optimization machinery to connect agentic \ValueAssign{} with joint admission and node placement. Its additive, lexicographic objective makes cache opportunity, tool readiness, and token footprint precedence explicit, and its asynchronous executor revalidates assignments against live state. Section~\ref{sec:exact-dp} and Appendix~\ref{sec:async-execution} specify snapshot optimality and the bounds for asynchronous application.

\paragraph{KV locality and agent scheduling.}
PagedAttention improves physical KV allocation and sharing, while SGLang's RadixAttention reuses prefixes within the serving runtime~\citep{vllm,sglang}. Preble co-optimizes distributed prefix reuse and computation balance; Mooncake combines a distributed KV store with disaggregated serving~\citep{preble,mooncake}. Dynamo and production gateways expose cache- and load-aware request scheduling~\citep{dynamo,vllmrouter,sglrouter}, and llm-d provides pluggable scoring components~\citep{llmd}. SMetric exploits session structure by balancing first-turn requests and assigning later turns to nodes with reusable cache, using a global cache tier to limit the cost of redistribution~\citep{smetric}. \Ours{} builds on these locality and load-balancing principles with joint optimization of assignments for waiting \Sessions{} under shared capacity. This joint view accounts for how one admission consumes capacity needed by other session--node assignments.

Program-level schedulers also preserve information across model calls. Autellix uses attained service and critical-path information to prioritize single-threaded and distributed programs~\citep{autellix}. ThunderAgent introduces the program abstraction, state-aware pausing, elapsed-time discounting of \Acting{} programs' effective token demand, and a global waiting queue~\citep{thunderagent}. \Ours{} adds node-specific cache values and \ReadyNext{} to the joint placement objective. Tool-gap cache management addresses a complementary decision: InferCept chooses among preservation, swapping, and recomputation by modeling resource waste~\citep{infercept}; Continuum selects a cache-retention TTL using historical tool-duration distributions, reconstruction cost, and returning programs' queueing cost, together with TTL-aware program ordering~\citep{continuum}. These systems control physical KV retention. \Ours{} coordinates scheduler-level admission opportunities, with predicted readiness distinguishing plans that have equal higher-precedence cache value. The serving engine controls the resulting cache residency.

\paragraph{Workflow, prediction, and tool execution.}
Workflow structure provides information beyond an isolated request. Parrot exposes inter-request dependencies through Semantic Variables, and Helium compiles workflow query plans for proactive reuse and cache-aware execution~\citep{parrot,helium}. KVFlow uses an Agent Step Graph and steps-to-execution to guide eviction and prefetching~\citep{kvflow}; CacheScout learns execution transitions online without a predefined graph~\citep{cachescout}. Their predictions guide context residency and prefetching; \ReadyNext{} supplies a binary near-term admission-opportunity signal derived from tool timing. SAGA combines workflow graphs, session-affinity batching, and work stealing~\citep{saga}. TOPAS also uses heuristic state searching: it scores prefix-residency and request-admission states using remaining workflow paths, movement costs, and downstream reuse, with bounded candidate search and greedy packing~\citep{topas}. Its task-dependent utility and engine-level residency decisions differ from our fixed additive value contract and exact multi-node scheduling problem.

Prediction also supports latency-oriented scheduling. ELIS predicts response length, while Kairos combines workflow-aware priority with memory-aware dispatch~\citep{elis,kairos}. Recent work separates already-ready turns from their release, using a mean--CVaR objective and an adaptive budget for committed inference work~\citep{turnrelease}. This addresses tail-aware release after dependencies complete; \Ours{} additionally evaluates near-term return opportunities during external tool execution. Complementary mechanisms change the execution timeline itself: Conveyor exposes partial tool execution, Sutradhara overlaps tool work and prefill through orchestrator--engine cooperation, and PASTE speculates on future tools while coordinating tool execution with returning model \Sessions{}~\citep{conveyor,sutradhara,paste}. SmoothAgent prepares transformed context on a lookahead stream, and Pie exposes programmable generation and KV handlers~\citep{smoothagent,pie}. \Ours{} operates in the scheduler, using timing evidence to coordinate admission and node placement.
\paragraph{Deployment and workloads.}
AgentSysBench documents substantial non-LLM latency and deployment-dependent resource demands~\citep{agentsysbench}. MARS explicitly coordinates GPU inference, CPU tool pressure, admission, and KV retention in heterogeneous serving~\citep{mars}. In our \Scene{} scenario, distributed executors manage tool CPU and memory, and their resource pressure reaches the GPU scheduler through tool-return timing. \Ours{} captures this interaction through readiness when allocating GPU admission capacity. OpenHands supports separate execution environments~\citep{openhands}. Coding and terminal workloads, together with InferenceX/AgentX trace replay, provide complementary evidence about task completion and serving behavior~\citep{miniswe,swebench,terminalbench,inferencex,agentx}, with workload-specific timing metrics defined in the evaluation.

%% file: references.bib
@article{bu2026tackling,
  title={Tackling the Data-Parallel Load Balancing Bottleneck in LLM Serving: Practical Online Routing at Scale},
  author={Bu, Tianci and Lyu, Yuan and Chen, Zixi and Song, Chendong and Liang, Hong and Gurung, Tsepten and Fan, Yuwei and Ye, Yinyu and Zhou, Zijie},
  journal={arXiv preprint arXiv:2605.06113},
  year={2026}
}

@article{chen2026online,
  title={Online Linear Programming for Multi-Objective Routing in LLM Serving},
  author={Chen, Zixi and Ye, Yinyu and Zhou, Zijie},
  journal={arXiv preprint arXiv:2607.03948},
  year={2026}
}

@article{li2026moe,
  title={MoE-SpAc: Efficient MoE Inference Based on Speculative Activation Utility in Heterogeneous Edge Scenarios},
  author={Li, Shuhuai and Lin, Jianghao and Ge, Dongdong and Ye, Yinyu},
  journal={arXiv preprint arXiv:2603.09983},
  year={2026}
}

@article{llm-core-mimo-scaling-rl,
  title={MiMo-V2. 6: Scaling reinforcement learning towards self-improvement},
  author={Qiao, Zongming and Hua, Ziyue and Ou, Zirui and Yue, Zihao and Jiang, Zihan and Huang, Zhuo and Chen, Zhiyang and Zheng, Zhixian and Xu, Zhipeng and Ma, Zhengrui and others},
  journal={arXiv preprint arXiv:2610.11959},
  year={2026}
}

@misc{glm47,
      title={GLM-4.5: Agentic, Reasoning, and Coding (ARC) Foundation Models},
      author={GLM-4.5 Team and Aohan Zeng and Xin Lv and Qinkai Zheng and Zhenyu Hou and Bin Chen and Chengxing Xie and Cunxiang Wang and Da Yin and Hao Zeng and Jiajie Zhang and Kedong Wang and Lucen Zhong and Mingdao Liu and Rui Lu and Shulin Cao and Xiaohan Zhang and Xuancheng Huang and Yao Wei and Yean Cheng and Yifan An and Yilin Niu and Yuanhao Wen and Yushi Bai and Zhengxiao Du and Zihan Wang and Zilin Zhu and Bohan Zhang and Bosi Wen and Bowen Wu and Bowen Xu and Can Huang and Casey Zhao and Changpeng Cai and Chao Yu and Chen Li and Chendi Ge and Chenghua Huang and Chenhui Zhang and Chenxi Xu and Chenzheng Zhu and Chuang Li and Congfeng Yin and Daoyan Lin and Dayong Yang and Dazhi Jiang and Ding Ai and Erle Zhu and Fei Wang and Gengzheng Pan and Guo Wang and Hailong Sun and Haitao Li and Haiyang Li and Haiyi Hu and Hanyu Zhang and Hao Peng and Hao Tai and Haoke Zhang and Haoran Wang and Haoyu Yang and He Liu and He Zhao and Hongwei Liu and Hongxi Yan and Huan Liu and Huilong Chen and Ji Li and Jiajing Zhao and Jiamin Ren and Jian Jiao and Jiani Zhao and Jianyang Yan and Jiaqi Wang and Jiayi Gui and Jiayue Zhao and Jie Liu and Jijie Li and Jing Li and Jing Lu and Jingsen Wang and Jingwei Yuan and Jingxuan Li and Jingzhao Du and Jinhua Du and Jinxin Liu and Junkai Zhi and Junli Gao and Ke Wang and Lekang Yang and Liang Xu and Lin Fan and Lindong Wu and Lintao Ding and Lu Wang and Man Zhang and Minghao Li and Minghuan Xu and Mingming Zhao and Mingshu Zhai and Pengfan Du and Qian Dong and Shangde Lei and Shangqing Tu and Shangtong Yang and Shaoyou Lu and Shijie Li and Shuang Li and Shuang-Li and Shuxun Yang and Sibo Yi and Tianshu Yu and Wei Tian and Weihan Wang and Wenbo Yu and Weng Lam Tam and Wenjie Liang and Wentao Liu and Xiao Wang and Xiaohan Jia and Xiaotao Gu and Xiaoying Ling and Xin Wang and Xing Fan and Xingru Pan and Xinyuan Zhang and Xinze Zhang and Xiuqing Fu and Xunkai Zhang and Yabo Xu and Yandong Wu and Yida Lu and Yidong Wang and Yilin Zhou and Yiming Pan and Ying Zhang and Yingli Wang and Yingru Li and Yinpei Su and Yipeng Geng and Yitong Zhu and Yongkun Yang and Yuhang Li and Yuhao Wu and Yujiang Li and Yunan Liu and Yunqing Wang and Yuntao Li and Yuxuan Zhang and Zezhen Liu and Zhen Yang and Zhengda Zhou and Zhongpei Qiao and Zhuoer Feng and Zhuorui Liu and Zichen Zhang and Zijun Yao and Zijun Yao and Zikang Wang and Ziqiang Liu and Ziwei Chai and Zixuan Li and Zuodong Zhao and Wenguang Chen and Jidong Zhai and Bin Xu and Minlie Huang and Hongning Wang and Juanzi Li and Yuxiao Dong and Jie Tang},
      year={2025},
      eprint={2508.06471},
      archivePrefix={arXiv},
      doi={10.48550/arXiv.2508.06471},
      primaryClass={cs.CL},
      url={https://arxiv.org/abs/2508.06471},
}

@misc{deepseek,
      title={DeepSeek-V4: Towards Highly Efficient Million-Token Context Intelligence}, 
      author={DeepSeek-AI and Anyi Xu and Bangcai Lin and Bing Xue and Bingxuan Wang and Bingzheng Xu and Bochao Wu and Bowei Zhang and Chaofan Lin and Chen Dong and Chenchen Ling and Chengda Lu and Chenggang Zhao and Chengqi Deng and Chengyu Hou and Chenhao Xu and Chenze Shao and Chong Ruan and Conner Sun and Damai Dai and Daya Guo and Dejian Yang and Deli Chen and Donghao Li and Dongjie Ji and Erhang Li and Fang Wei and Fangyun Lin and Fangzhou Yuan and Feiyu Xia and Fucong Dai and Guangbo Hao and Guanting Chen and Guoai Cao and Guolai Meng and Guowei Li and Han Yu and Han Zhang and Hanwei Xu and Hao Li and Haofen Liang and Haoling Zhang and Haoming Luo and Haoran Wei and Haotian Yuan and Haowei Zhang and Haowen Luo and Haoyu Chen and Haozhe Ji and Hengqing Zhang and Honghui Ding and Hongxuan Tang and Huanqi Cao and Huazuo Gao and Hui Qu and Hui Zeng and J Yang and JQ Zhu and Jia Luo and Jia Song and Jia Yu and Jialiang Huang and Jialu Cai and Jian Liang and Jiangting Zhou and Jiasheng Ye and Jiashi Li and Jiaxin Xu and Jiewen Hu and Jieyu Yang and Jin Chen and Jin Yan and Jingchang Chen and Jingli Zhou and Jingting Xiang and Jingyang Yuan and Jingyuan Cheng and Jingzi Zhou and Jinhua Zhu and Jiping Yu and Joseph Sun and Jun Ran and Junguang Jiang and Junjie Qiu and Junlong Li and Junmin Zheng and Junxiao Song and Kai Dong and Kaige Gao and Kang Guan and Kexing Zhou and Kezhao Huang and Kuai Yu and Lean Wang and Lecong Zhang and Lei Wang and Leyi Xia and Li Zhang and Liang Zhao and Lihua Guo and Lingxiao Luo and Linwang Ma and Linyan Zhu and Litong Wang and Liyu Cai and Liyue Zhang and Longhao Chen and MS Di and MY Xu and Max Mei and Miaojun Wang and Mingchuan Zhang and Minghua Zhang and Minghui Tang and Mingming Li and Mingxu Zhou and Minmin Han and Ning Wang and Panpan Huang and Panpan Wang and Peixin Cong and Peiyi Wang and Peng Zhang and Qiancheng Wang and Qihao Zhu and Qingyang Li and Qinyu Chen and Qiushi Du and Qiwei Jiang and Rui Tian and Ruifan Xu and Ruijie Lu and Ruiling Xu and Ruiqi Ge and Ruisong Zhang and Ruizhe Pan and Runji Wang and Runqian Chen and Runqiu Yin and Runxin Xu and Ruomeng Shen and Ruoyu Zhang and Ruyi Chen and SH Liu and Shanghao Lu and Shangmian Sun and Shangyan Zhou and Shanhuang Chen and Shaofei Cai and Shaoheng Nie and Shaoqing Wu and Shaoyuan Chen and Shengding Hu and Shengyu Liu and Shiqiang Hu and Shirong Ma and Shiyu Wang and Shuiping Yu and Shunfeng Zhou and Shuting Pan and Shuying Yu and Songyang Zhou and Tao Ni and Tao Yun and Tian Jin and Tian Pei and Tian Ye and Tianle Lin and Tianran Ji and Tianyi Cui and Tianyuan Yue and Tingting Yu and Tun Wang and W Zhang and WL Xiao and Wangding Zeng and Wei An and Weilin Zhao and Wen Liu and Wenfeng Liang and Wenjie Pang and Wenjing Luo and Wenjing Yao and Wenjun Gao and Wenkai Yang and Wenlve Huang and Wenqing Hou and Wentao Zhang and Wenting Ma and Xi Gao and Xiang He and Xiangwen Wang and Xianzu Wang and Xiao Bi and Xiaodong Liu and Xiaohan Wang and Xiaokang Chen and Xiaokang Zhang and Xiaotao Nie and Xiaowen Sun and Xiaoxiang Wang and Xin Cheng and Xin Liu and Xin Xie and Xingchao Liu and Xingchen Liu and Xingkai Yu and Xingyou Li and Xinyu Yang and Xinyu Zhang and Xu Chen and Xuanyu Wang and Xuecheng Su and Xueyin Chen and Xuheng Lin and Xuwei Fu and YC Yan and YQ Wang and YW Ma and Yanfeng Luo and Yang Zhang and Yanhong Xu and Yanru Ma and Yanwen Huang and Yao Li and Yao Li and Yao Xu and Yao Zhao and Yaofeng Sun and Yaohui Wang and Yi Qian and Yi Shao and Yi Yu and Yichao Zhang and Yifan Ding and Yifan Shi and Yijia Wu and Yiliang Xiong and Yiling Ma and Ying He and Ying Tang and Ying Zhou and Yingjia Luo and Yinmin Zhong and Yishi Piao and Yisong Wang and Yixiang Zhang and Yixiao Chen and Yixuan Tan and Yixuan Wei and Yiyang Ma and Yiyuan Liu and Yonglun Yang and Yongqiang Guo and Yongtong Wu and Yu Wu and YuKun Li and Yuan Cheng and Yuan Ou and Yuanfan Xu and Yuanhao Li and Yuduan Wang and Yuehan Yang and Yuer Xu and Yuhan Wu and Yuhao Meng and Yuheng Zou and Yukun Zha and Yunfan Xiong and Yupeng Chen and Yuping Lin and Yuqian Cao and Yuqian Wang and Yushun Zhang and Yuting Yan and Yutong Lin and Yuxian Gu and Yuxiang Luo and Yuxiang You and Yuxuan Liu and Yuxuan Zhou and Yuyang Zhou and Yuzhen Huang and ZF Wu and Zehao Wang and Zehua Zhao and Zehui Ren and Zekai Zhang and Zhangli Sha and Zhe Fu and Zhe Ju and Zhean Xu and Zhenda Xie and Zhengyan Zhang and Zheren Gao and Zhewen Hao and Zhibin Gou and Zhicheng Ma and Zhigang Yan and Zhihong Shao and Zhixian Huang and Zhixuan Chen and Zhiyu Wu and Zhizhou Ren and Zhongyu Wu and Zhuoshu Li and Zhuping Zhang and Zian Xu and Zihao Wang and Zihua Qu and Zihui Gu and Zijia Zhu and Zilin Li and Zipeng Zhang and Ziwei Xie and Ziyi Gao and Ziyi Wan and Zizheng Pan and Zongqing Yao},
      year={2026},
      eprint={2606.19348},
      archivePrefix={arXiv},
      primaryClass={cs.CL},
      url={https://arxiv.org/abs/2606.19348}, 
}

@misc{qwen3,
      title={Qwen3 Technical Report}, 
      author={An Yang and Anfeng Li and Baosong Yang and Beichen Zhang and Binyuan Hui and Bo Zheng and Bowen Yu and Chang Gao and Chengen Huang and Chenxu Lv and Chujie Zheng and Dayiheng Liu and Fan Zhou and Fei Huang and Feng Hu and Hao Ge and Haoran Wei and Huan Lin and Jialong Tang and Jian Yang and Jianhong Tu and Jianwei Zhang and Jianxin Yang and Jiaxi Yang and Jing Zhou and Jingren Zhou and Junyang Lin and Kai Dang and Keqin Bao and Kexin Yang and Le Yu and Lianghao Deng and Mei Li and Mingfeng Xue and Mingze Li and Pei Zhang and Peng Wang and Qin Zhu and Rui Men and Ruize Gao and Shixuan Liu and Shuang Luo and Tianhao Li and Tianyi Tang and Wenbiao Yin and Xingzhang Ren and Xinyu Wang and Xinyu Zhang and Xuancheng Ren and Yang Fan and Yang Su and Yichang Zhang and Yinger Zhang and Yu Wan and Yuqiong Liu and Zekun Wang and Zeyu Cui and Zhenru Zhang and Zhipeng Zhou and Zihan Qiu},
      year={2025},
      eprint={2505.09388},
      archivePrefix={arXiv},
      primaryClass={cs.CL},
      url={https://arxiv.org/abs/2505.09388}, 
}

@misc{dsec,
      title={DeepSeek Elastic Compute (DSec): A Sandbox Infrastructure for Effective Agentic Training at Scale}, 
      author={Jialiang Huang and Hongxuan Tang and Jingchang Chen and Yuxuan Liu and Yixiao Chen and Yuan Cheng and Yi Tao and Jingli Zhou and Yupeng Chen and Haoyu Chen and Jiarui Wang and Shengkai Lin and Chuqi Zhang and Bryan Lee Teng and Lian Guo and Zhe Fu and Wenjun Gao and Yisong Wang and Liang Zhao and Zehao Wang and Ziwei Xie and Yongqiang Guo and Peixin Cong and Ziyi Gao and Shuiping Yu and Hanwei Xu and Zuofan Wu and Zhizhou Ren and Yuyang Zhou and Bowei Zhang and Zhihuan Huang and Qihao Zhu and Lei Wang and Tianle Lin and Han Yu and Jiewen Hu and Dejian Yang and Shuo Yang and Shanghao Lu and Shaoyuan Chen and Junjie Qiu and Zhangli Sha and Yinmin Zhong and Yongtong Wu and Shiyu Wang and Wei Liu and Bingzheng Xu and Longhao Chen and Qiushi Du and Yuzhen Huang and Shirong Ma and Yaohui Wang and Mingshu Chen and Tongrui Xiong and Y. C. Yan and Haowen Luo and Haofen Liang and Xiaokang Zhang and Weihao Zeng and Runxin Xu and Peiyi Wang and Jinhua Zhu and Ruoyu Zhang and Wenkai Yang and Qi Tang and Jiping Yu and Tian Ye and Ruizhe Pan and Honghui Ding and Xiaodong Liu and Lingxiao Luo and Zhihong Shao and Yuhan Wu and Jibai Lu and Wen Liu and Haoling Zhang and Jingcheng Hu and Yaoyang Ye and Chaofan Lin and Zhaochen Zhang and Jianan Tong and Hengxu Wu and Zhihao Li and Yicheng Wang and Luyao Wang and Yuzhuo Bai and Lingyue Fu and Ruifan Xu and Y. Z. Wang and Zonglin Li and Mingqi Wei and Haiyang Shen and Chengyuan Zhang and Chao Jin and Zili Zhang and R. H. Yang and Xinbo Xu and Jian Zhou and Ruidong Zhu and Yuzhe Guo and Zelun Pan and Shaoheng Nie and Erhang Li and Shuhan Lin and Zheng Liu and Anshuo Chen and Zilong Lyu and Sinuo Cao and Rui Yu and Chuhao Wang and Junyi Guo and Junxiao Song and Kaifeng Chen and Menghao Ye and Junxian Li and Di Wu and Haiyang Ma and Yilun Wang and Haoran Yang and Yizai Cai and Shichun Liu and Yiping Wang and Junbo Sun and Shicheng Xu and Xiao Bi and Ying He and Yichao Zhang and Mingxing Zhang and Liyue Zhang and Panpan Huang and Wenfeng Liang},
      year={2026},
      eprint={2609.22978},
      archivePrefix={arXiv},
      primaryClass={cs.DC},
      url={https://arxiv.org/abs/2609.22978}, 
}

@article{bernstein,
  title={Empirical bernstein bounds and sample variance penalization},
  author={Maurer, Andreas and Pontil, Massimiliano},
  journal={arXiv preprint arXiv:0907.3740},
  year={2009}
}

@article{hexagent,
  title={HexAGenT: Efficient Agentic LLM Serving via Workflow-and Heterogeneity-Aware Scheduling},
  author={Peng, You and Jiang, Youhe and Li, Wenshuang and Xu, Xu and Zhou, Ke and Jiang, Jiawei and Wang, Chen and Yuan, Binhang},
  journal={arXiv preprint arXiv:2605.16637},
  year={2026}
}

@misc{anthropic_tooluse,
  title={How Tool Use Works},
  author={{Anthropic}},
  year={2026},
  howpublished={\url{https://platform.claude.com/docs/en/agents-and-tools/tool-use/how-tool-use-works}},
  note={Official documentation; accessed September 11, 2026}
}

@misc{openai_computeruse,
  title={Computer Use},
  author={{OpenAI}},
  year={2026},
  howpublished={\url{https://developers.openai.com/api/docs/guides/tools-computer-use}},
  note={Official documentation; accessed September 11, 2026}
}

@article{kairos_power,
  title={KAIROS: Stateful, Context-Aware Power-Efficient Agentic Inference Serving},
  author={Yuan, Yichao and Chowdhury, Mosharaf and Talati, Nishil},
  journal={arXiv preprint arXiv:2604.16682},
  year={2026}
}

@misc{vllmrouter,
  title={{vLLM Production Stack}},
  author={{vLLM Production Stack Team}},
  year={n.d.},
  howpublished={\url{https://docs.vllm.ai/projects/production-stack/en/latest/}},
  note={Technical documentation; accessed September 7, 2026}
}

@misc{claude_monitor,
  title={{Claude Code} Tools Reference: Monitor Tool},
  author={{Anthropic}},
  year={2026},
  howpublished={\url{https://code.claude.com/docs/en/tools-reference}},
  note={Accessed September 8, 2026}
}

@misc{opencode_timeout,
  title={Improve Bash Timeout Retry Hint},
  author={{OpenCode Contributors}},
  year={2026},
  howpublished={\url{https://github.com/anomalyco/opencode/pull/22390}},
  note={Pull request 22390, merged April 14, 2026; timeout-termination metadata}
}

@misc{llmd,
  title={Intelligent Inference Scheduling with {llm-d}},
  author={Guy, Nili and Bortnikov, Vita and {Lev Ran}, Etai and Shaw, Robert and Coleman, Clayton},
  year={2025},
  howpublished={\url{https://llm-d.ai/blog/intelligent-inference-scheduling-with-llm-d}},
  note={Published September 3, 2025}
}

@misc{sglrouter,
  title={{SGLang Model Gateway}},
  author={{SGLang Team}},
  year={n.d.},
  howpublished={\url{https://docs.sglang.io/docs/advanced_features/sgl_model_gateway}},
  note={Technical documentation; accessed September 7, 2026}
}

@inproceedings{vllm,
  title={Efficient Memory Management for Large Language Model Serving with PagedAttention},
  author={Kwon, Woosuk and Li, Zhuohan and Zhuang, Siyuan and Sheng, Ying and Zheng, Lianmin and Yu, Cody Hao and Gonzalez, Joseph E. and Zhang, Hao and Stoica, Ion},
  booktitle={Proceedings of the ACM SIGOPS Symposium on Operating Systems Principles},
  year={2023},
  url={https://arxiv.org/abs/2309.06180v1}
}

@inproceedings{sglang,
  title={{SGLang}: Efficient Execution of Structured Language Model Programs},
  author={Zheng, Lianmin and Yin, Liangsheng and Xie, Zhiqiang and Sun, Chuyue and Huang, Jeff and Yu, Cody Hao and Cao, Shiyi and Kozyrakis, Christos and Stoica, Ion and Gonzalez, Joseph E. and Barrett, Clark and Sheng, Ying},
  booktitle={Advances in Neural Information Processing Systems},
  volume={37},
  pages={62557--62583},
  year={2024},
  doi={10.52202/079017-2000},
  url={https://proceedings.neurips.cc/paper_files/paper/2024/hash/724be4472168f31ba1c9ac630f15dec8-Abstract-Conference.html}
}

@inproceedings{preble,
  title={{Preble}: Efficient Distributed Prompt Scheduling for {LLM} Serving},
  author={Srivatsa, Vikranth and He, Zijian and Abhyankar, Reyna and Li, Dongming and Zhang, Yiying},
  booktitle={International Conference on Learning Representations},
  year={2025},
  url={https://proceedings.iclr.cc/paper_files/paper/2025/hash/5bc342f48de8264779952fac378f96dc-Abstract-Conference.html}
}

@inproceedings{mooncake,
  title={{Mooncake}: Trading More Storage for Less Computation---A {KVCache}-centric Architecture for Serving {LLM} Chatbot},
  author={Qin, Ruoyu and Li, Zheming and He, Weiran and Cui, Jialei and Ren, Feng and Zhang, Mingxing and Wu, Yongwei and Zheng, Weimin and Xu, Xinran},
  booktitle={23rd USENIX Conference on File and Storage Technologies},
  pages={155--170},
  publisher={USENIX Association},
  year={2025},
  url={https://www.usenix.org/conference/fast25/presentation/qin}
}

@misc{dynamo,
  title={NVIDIA Dynamo: A Distributed Inference Framework},
  author={{NVIDIA}},
  year={2025},
  howpublished={\url{https://docs.nvidia.com/dynamo/}},
  note={Technical documentation; scheduler code audited in the literature ledger at commit 1e3478dfe09bdd305d7c58f412fda2dfe5bec7ed}
}

@article{helium,
  title={Efficient LLM Serving for Agentic Workflows: A Data Systems Perspective},
  author={Wadlom, Noppanat and Shen, Junyi and Lu, Yao},
  journal={arXiv preprint arXiv:2603.16104},
  year={2026},
  url={https://arxiv.org/abs/2603.16104}
}

@article{smoothagent,
  title={SmoothAgent: Efficient Long-Horizon LLM-Based Agent Serving with Lookahead Context Engineering},
  author={Pan, Zaifeng and Wang, Qianxu and Hu, Zhengding and Chen, Chang and Guan, Yue and Zhou, Yanbo and Swanson, Steven and Ding, Yufei},
  journal={arXiv preprint arXiv:2607.00151},
  year={2026}
}

@article{elis,
  title={{ELIS}: Efficient {LLM} Iterative Scheduling System with Response Length Predictor},
  author={Choi, Seungbeom and Goo, Jeonghoe and Jeon, Eunjoo and Yang, Mingyu and Jang, Minsung},
  journal={arXiv preprint arXiv:2505.09142},
  year={2025},
  url={https://arxiv.org/abs/2505.09142}
}

@article{kairos,
  title={{Kairos}: Low-latency Multi-Agent Serving with Shared {LLMs} and Excessive Loads in the Public Cloud},
  author={Chen, Jinyuan and Shi, Jiuchen and Chen, Quan and Guo, Minyi},
  journal={arXiv preprint arXiv:2508.06948},
  year={2025},
  url={https://arxiv.org/abs/2508.06948}
}

@article{agentsysbench,
  title={From LLM Inference to Agentic Workloads: Characterization and Implications for Serving Systems},
  author={Chang, Chaokun and Zhou, Yukun and Fu, Kaihua and An, Dakai and Feng, Tianyu and Lu, Hanfeng and Yao, Sheng and Guo, Pu and Yu, Yinghao and Shan, Yizhou and Li, Bo and Yuan, Binhang and Wang, Wei},
  journal={arXiv preprint arXiv:2608.15127},
  year={2026},
  url={https://arxiv.org/abs/2608.15127v1}
}

@book{knapsack,
  title={Knapsack Problems},
  author={Kellerer, Hans and Pferschy, Ulrich and Pisinger, David},
  publisher={Springer},
  year={2004},
  doi={10.1007/978-3-540-24777-7}
}

@article{shmoystardos,
  title={An approximation algorithm for the generalized assignment problem},
  author={Shmoys, David B. and Tardos, Eva},
  journal={Mathematical Programming},
  volume={62},
  pages={461--474},
  year={1993},
  doi={10.1007/BF01585178}
}

@article{conveyor,
  title={{Conveyor}: Efficient Tool-aware {LLM} Serving with Tool Partial Execution},
  author={Xu, Yechen and Kong, Xinhao and Chen, Tingjun and Zhuo, Danyang},
  journal={arXiv preprint arXiv:2406.00059},
  year={2024},
  url={https://arxiv.org/abs/2406.00059}
}

@article{thunderagent,
  title={{ThunderAgent}: A Simple, Fast and Program-Aware Agentic Inference System},
  author={Kang, Hao and Li, Ziyang and Xu, Weili and Yang, Xinyu and Chen, Yinfang and Wang, Junxiong and Chen, Beidi and Krishna, Tushar and Xu, Chenfeng and Arora, Simran},
  journal={arXiv preprint arXiv:2602.13692},
  year={2026},
  url={https://arxiv.org/abs/2602.13692v3},
  note={Version 3, June 30, 2026}
}

@book{bellman,
  title={Dynamic Programming},
  author={Bellman, Richard},
  publisher={Princeton University Press},
  year={1957}
}

@article{pisinger,
  title={An exact algorithm for large multiple knapsack problems},
  author={Pisinger, David},
  journal={European Journal of Operational Research},
  volume={114},
  number={3},
  pages={528--541},
  year={1999},
  doi={10.1016/S0377-2217(98)00120-9}
}

@article{chekurikhanna,
  title={A Polynomial Time Approximation Scheme for the Multiple Knapsack Problem},
  author={Chekuri, Chandra and Khanna, Sanjeev},
  journal={SIAM Journal on Computing},
  volume={35},
  number={3},
  pages={713--728},
  year={2005},
  doi={10.1137/S0097539700382820}
}

@article{lisunye,
  title={Simple and Fast Algorithm for Binary Integer and Online Linear Programming},
  author={Li, Xiaocheng and Sun, Chunlin and Ye, Yinyu},
  journal={arXiv preprint arXiv:2003.02513},
  year={2020},
  url={https://arxiv.org/abs/2003.02513v3}
}

@article{cachescout,
  title={Learning Agent Execution for {KV}-Cache Management in Agentic Serving},
  author={Zhang, Rui and Kim, Chaeeun and Feng, Shaoting and Du, Kuntai and Liu, Yuhan and Zhong, Yi and Ching, Cheng-Wei and Jiang, Junchen and Hu, Liting},
  journal={arXiv preprint arXiv:2608.14624},
  year={2026},
  url={https://arxiv.org/abs/2608.14624v1}
}

@article{tokencake,
  title={{Tokencake}: A {KV}-Cache-centric Serving Framework for {LLM}-based Multi-Agent Applications},
  author={Bian, Zhuohang and Wu, Feiyang and Ma, Teng and Zhuo, Youwei},
  journal={arXiv preprint arXiv:2510.18586},
  year={2025},
  url={https://arxiv.org/abs/2510.18586v2}
}

@article{smetric,
  title={{SMetric}: Rethink {LLM} Scheduling for Serving Agents with Balanced Session-centric Scheduling},
  author={Wang, Jiahao and Lin, Kaizhan and Zhang, Kaixi and Han, Jinbo and Wei, Xingda and Shen, Sijie and Fang, Chenguang and Yu, Wenyuan and Chen, Rong and Chen, Haibo},
  journal={arXiv preprint arXiv:2607.08565},
  year={2026},
  url={https://arxiv.org/abs/2607.08565v1}
}

@article{sutradhara,
  title={{Sutradhara}: An Intelligent Orchestrator-Engine Co-design for Tool-based Agentic Inference},
  author={Biswas, Anish and Goel, Kanishk and {Srivarshinee S} and Mohan, Jayashree and Khare, Alind and Parayil, Anjaly and Ramjee, Ramachandran and Bansal, Chetan},
  journal={arXiv preprint arXiv:2601.12967},
  year={2026},
  url={https://arxiv.org/abs/2601.12967v3}
}

@article{paste,
  title={Parallelizing Tool Execution and {LLM} Generation for Low-Latency Agent Serving},
  author={Sui, Yifan and Zhao, Han and Ma, Rui and He, Zhiyuan and Wang, Hao and Li, Jianxun and Xu, Kaiqiang and Chen, Kai and Yang, Yuqing},
  journal={arXiv preprint arXiv:2603.18897},
  year={2026},
  url={https://arxiv.org/abs/2603.18897v3}
}

@article{continuum,
  title={{Continuum}: Efficient and Robust Multi-Turn {LLM} Agent Scheduling with {KV} Cache Time-to-Live},
  author={Li, Hanchen and He, Runyuan and Mang, Qiuyang and Zhang, Qizheng and Mao, Huanzhi and Chen, Xiaokun and Zhou, Hangrui and Zhang, Huanchen and Cheung, Alvin and Gonzalez, Joseph and Stoica, Ion},
  journal={arXiv preprint arXiv:2511.02230},
  year={2025},
  url={https://arxiv.org/abs/2511.02230v7},
  note={Version 7, September 8, 2026}
}

@inproceedings{infercept,
  title={{InferCept}: Efficient Intercept Support for Augmented Large Language Model Inference},
  author={Abhyankar, Reyna and He, Zijian and Srivatsa, Vikranth and Zhang, Hao and Zhang, Yiying},
  booktitle={Proceedings of the 41st International Conference on Machine Learning},
  series={Proceedings of Machine Learning Research},
  volume={235},
  pages={81--95},
  publisher={PMLR},
  year={2024},
  url={https://proceedings.mlr.press/v235/abhyankar24a.html}
}

@inproceedings{parrot,
  title={{Parrot}: Efficient Serving of {LLM}-based Applications with Semantic Variable},
  author={Lin, Chaofan and Han, Zhenhua and Zhang, Chengruidong and Yang, Yuqing and Yang, Fan and Chen, Chen and Qiu, Lili},
  booktitle={18th USENIX Symposium on Operating Systems Design and Implementation},
  pages={929--945},
  publisher={USENIX Association},
  year={2024},
  url={https://www.usenix.org/conference/osdi24/presentation/lin-chaofan}
}

@inproceedings{kvflow,
  title={{KVFlow}: Efficient Prefix Caching for Accelerating {LLM}-Based Multi-Agent Workflows},
  author={Pan, Zaifeng and Patel, Ajjkumar Dahyalal and Shen, Yipeng and Hu, Zhengding and Guan, Yue and Li, Wan-Lu and Qin, Lianhui and Wang, Yida and Ding, Yufei},
  booktitle={Advances in Neural Information Processing Systems},
  volume={38},
  pages={126246--126265},
  year={2025},
  doi={10.52202/085713-4208},
  url={https://papers.neurips.cc/paper_files/paper/2025/hash/b7971d31a7d5eb0f1eed2f8f6f368195-Abstract-Conference.html}
}

@inproceedings{pie,
  title={{Pie}: A Programmable Serving System for Emerging {LLM} Applications},
  author={Gim, In and Ma, Zhiyao and Lee, Seung-seob and Zhong, Lin},
  booktitle={Proceedings of the ACM SIGOPS Symposium on Operating Systems Principles},
  year={2025},
  doi={10.1145/3731569.3764814},
  url={https://arxiv.org/abs/2510.24051v1}
}

@article{mars,
  title={{MARS}: Efficient, Adaptive Co-Scheduling for Heterogeneous Agentic Systems},
  author={Wang, Yifei and Ye, Hancheng and Xu, Yechen and Guo, Cong and Wei, Chiyue and Wang, Qinsi and Li, Dongting and Chen, Tingjun and Li, Hai {``Helen''} and Zhuo, Danyang and Chen, Yiran},
  journal={arXiv preprint arXiv:2604.26963},
  year={2026},
  url={https://arxiv.org/abs/2604.26963v2},
  note={Version 2, June 15, 2026}
}

@article{saga,
  title={{SAGA}: Workflow-Atomic Scheduling for {AI} Agent Inference on {GPU} Clusters},
  author={Guo, Dongxin and Wu, Jikun and Yiu, Siu Ming},
  journal={arXiv preprint arXiv:2605.00528},
  year={2026},
  url={https://arxiv.org/abs/2605.00528v2},
  note={Version 2, June 19, 2026}
}

@article{topas,
  title={{TOPAS}: Workflow-Aware Prefix-State Scheduling for Multi-Agent {LLM} Serving},
  author={Ni, Hongqiu and Tian, Han and Zhang, Chi and Li, Guopeng and Tan, Haisheng},
  journal={arXiv preprint arXiv:2608.25523},
  year={2026},
  url={https://arxiv.org/abs/2608.25523v1},
  note={Version 1, August 26, 2026}
}

@article{turnrelease,
  title={Decoupling Readiness from Release for Tail-Aware Scheduling of Agentic {LLM} Workflows},
  author={Feng, Bochao and Li, Jianjiang and Wang, Haojie and Qiao, Lin and Li, Yinghui and Yan, Yukun and Zhai, Jidong},
  journal={arXiv preprint arXiv:2609.10964},
  year={2026},
  url={https://arxiv.org/abs/2609.10964v1},
  note={Version 1, September 10, 2026}
}

@article{autellix,
  title={{Autellix}: An Efficient Serving Engine for {LLM} Agents as General Programs},
  author={Luo, Michael and Shi, Xiaoxiang and Cai, Colin and Zhang, Tianjun and Wong, Justin and Wang, Yichuan and Wang, Chi and Huang, Yanping and Chen, Zhifeng and Gonzalez, Joseph E. and Stoica, Ion},
  journal={arXiv preprint arXiv:2502.13965},
  year={2025},
  url={https://arxiv.org/abs/2502.13965v1}
}

@misc{miniswe,
  title={{mini-SWE-agent}},
  author={{mini-SWE-agent Contributors}},
  year={2026},
  howpublished={\url{https://mini-swe-agent.com/latest/}},
  note={Official documentation; accessed September 8, 2026}
}

@inproceedings{swebench,
  title={{SWE}-bench: Can Language Models Resolve Real-world {Github} Issues?},
  author={Jimenez, Carlos E. and Yang, John and Wettig, Alexander and Yao, Shunyu and Pei, Kexin and Press, Ofir and Narasimhan, Karthik R.},
  booktitle={International Conference on Learning Representations},
  year={2024},
  url={https://openreview.net/forum?id=VTF8yNQM66}
}

@inproceedings{terminalbench,
  title={Terminal-bench: Benchmarking agents on hard, realistic tasks in command line interfaces},
  author={Merrill, Mike and Shaw, Alexander and Carlini, Nicholas and Li, Boxuan and Raj, Harsh and Bercovich, Ivan and Shi, Lin and Shin, Jeong and Walshe, Thomas and Buchanan, E Kelly and others},
  booktitle={International Conference on Learning Representations},
  volume={2026},
  pages={40903--40986},
  year={2026}
}

@misc{inferencex,
  title={{InferenceX}: Open Source Continuous Inference Benchmark Research Platform},
  author={{SemiAnalysis}},
  year={2026},
  howpublished={\url{https://github.com/SemiAnalysisAI/InferenceX}},
  note={Official repository; accessed September 8, 2026}
}

@misc{agentx,
  title={{AgentX MVP}: Multi-turn Coding-agent Trace Replay},
  author={{SemiAnalysis}},
  year={2026},
  howpublished={\url{https://github.com/SemiAnalysisAI/agentx-harness/blob/master/docs/tutorials/agentx-mvp.md}},
  note={Official documentation; accessed September 8, 2026}
}

@inproceedings{openhands,
  title={Openhands: An open platform for ai software developers as generalist agents},
  author={Wang, Xingyao and Li, Boxuan and Song, Yufan and Xu, Frank F and Tang, Xiangru and Zhuge, Mingchen and Pan, Jiayi and Song, Yueqi and Li, Bowen and Singh, Jaskirat and others},
  booktitle={International Conference on Learning Representations},
  volume={2025},
  pages={65882--65919},
  year={2025}
}

@inproceedings{agentix,
  title={{Agentix}: An Efficient Serving Engine for {LLM} Agents as General Programs},
  author={Luo, Michael and Shi, Xiaoxiang and Cai, Colin and Zhang, Tianjun and Wong, Justin and Wang, Yichuan and Wang, Chi and Huang, Yanping and Chen, Zhifeng and Gonzalez, Joseph E. and Stoica, Ion},
  booktitle={23rd USENIX Symposium on Networked Systems Design and Implementation (NSDI 26)},
  year={2026},
  pages={2443--2459},
  publisher={USENIX Association},
  url={https://www.usenix.org/conference/nsdi26/presentation/luo}
}
